\documentclass[twocolumn]{revtex4}

\usepackage{graphicx}
\usepackage[dvips]{color}
\usepackage{epsfig}
\usepackage{amssymb}
\usepackage{amsmath}
\usepackage{bm}

\begin{document}

\bibliographystyle{apsrev}
\title{Nutrient Shielding in Clusters of  Cells}
\author{Maxim O.~Lavrentovich${}^1$ }
\author{John H.~Koschwanez${}^{2,3}$}
\author{David R. Nelson${}^{1,2,3}$}
\affiliation{${}^{1}$Department of Physics}
\affiliation{${}^{2}$Department of Molecular and Cellular Biology}
\affiliation{${}^{3}$FAS Center for
Systems Biology, Harvard University, Cambridge, Massachusetts 02138, USA}
\begin{abstract}

Cellular nutrient consumption is influenced by both the nutrient uptake kinetics of an individual cell and the cells'  spatial  arrangement.    Large cell clusters or colonies have inhibited growth at the cluster's center  due to the shielding of nutrients by the cells closer to the surface.  We develop an effective medium theory that predicts a  thickness $\ell$ of the outer shell of cells in the cluster that receives enough nutrient to grow.  The cells are treated as partially absorbing identical spherical nutrient sinks, and  we identify a dimensionless parameter $\nu$ that characterizes the absorption strength of each cell.  The parameter $\nu$ can vary over many orders of magnitude between different cell types, ranging from bacteria and yeast to human tissue. The thickness $\ell$ decreases with increasing  $\nu$, increasing cell volume fraction $\phi$, and decreasing ambient nutrient concentration $\psi_{\infty}$.  The theoretical results are compared with numerical simulations and experiments.  In the latter studies, colonies of budding yeast, \textit{Saccharomyces cerevisiae},  are grown on glucose media and imaged under a confocal microscope. We measure the growth inside the colonies via a fluorescent protein reporter and compare the experimental and theoretical results for  the  thickness  $\ell$.

\end{abstract}

\date{\today}

\maketitle

\section{\label{SIntro}Introduction}

 Nutrient uptake is essential for all life and has been studied in a variety of model organisms; however, the physical mechanisms involved in the uptake are not yet well understood.  A cell commonly takes up nutrients from its surrounding medium via facilitated diffusion or active transport:   Specialized transporters on the cell surface move the nutrient down (facilitated diffusion) or up (active transport) a concentration gradient  from the ambient medium  into the cell.  The cell's absorption of nutrients in its immediate vicinity sets up and maintains a concentration gradient outside the cell that gradually depletes the nutrient from the surrounding medium.
  This is the method used by budding yeast cells for acquiring glucose, for example \cite{difftrans,transyeast, meijer}.   

\begin{figure}
\includegraphics[height=2in]{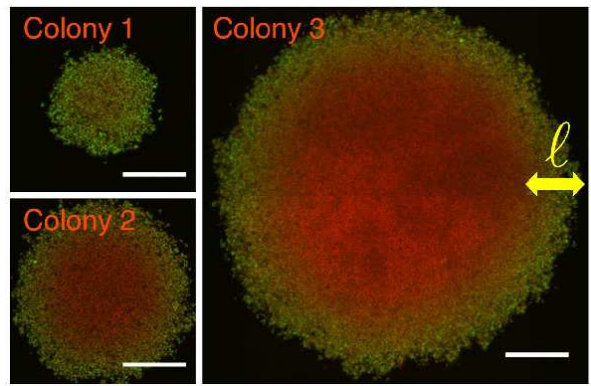}
\caption{\label{fconfocal} (Color online) Confocal microscope images of cross-sections through the bottom of three budding yeast colonies (scale bars represent 100 $\mu$m).       The red (darker shade) color is the constitutive expression of a protein,  whose level is largely independent of growth rate, in all cells and the green (lighter shade) color is ribosomal protein expression, indicating growth. Colony 1: 0.5 mM glucose, 43 h after inoculation; Colony 2: 1.5 mM glucose, 47 h after inoculation; Colony 3: 4.5 mM glucose, 56 h after inoculation. Small colonies such as Colony 1 receive enough nutrients for all cells to grow.  Colonies 2 and 3 are larger and growth occurs only in an outer shell of thickness $\ell$ (illustrated for Colony 3).  The red (darkly shaded) cells in the interior are shielded from the nutrients.    We assume the colonies have a spherical cap shape (see Sec.~\ref{SExperiments} for a discussion of the shape and experimental details).}
\end{figure}

 Nutrient consumption via facilitated diffusion is ubiquitous in nature and occurs in a variety of contexts, including oxygen consumption by human tissue cells, calcium uptake by intestinal cells, bacterial absorption of various sugars, etc \cite{oxygenreview,calciumreview,sugarreview}.   In many of these examples, cells grow in clusters and shield each other from the available nutrients.  In Fig.~\ref{fconfocal}, we see confocal microscope images of yeast colony cross sections in which yeast cells marked green (light shade) are growing due to an abundance of glucose.  Nutrient (glucose) shielding in larger colonies (Colonies 2 and 3) prevents cell growth  in the colony interiors (red (darkly shaded) regions in Fig.~\ref{fconfocal}) and only an outer shell of thickness $\ell$ is able to grow (illustrated for Colony 3 in Fig.~\ref{fconfocal}).   

 Typical bacterial and yeast cell colonies (e.g., those in Fig.~\ref{fconfocal}) are dense cell clusters  with large cell packing fractions $\phi \gtrsim 0.5$ \cite{bacphi,yeastphi}.  In this paper, we characterize cells packed in spherical clusters at various values of  $\phi$.  The shape of colonies in the experiments shown in Fig.~\ref{fconfocal} can be approximated by a dome-shaped section of a sphere, as follows from the observation that yeast colonies growing on flat surfaces can be described by a contact angle \cite{colonyshape}.     Our theoretical calculations for nutrient shielding by complete spheres should apply whenever the penetration depth $\ell$ is small compared to the dome height or when the domes are approximately hemispherical.

 Dilute clusters with $\phi \lesssim 0.1$ can be constructed artificially, by suspending microbial cells in a gelatinous matrix.  Immobilized microbial cells have a very wide range of industrial and environmental applications (see Ref.~\cite{embed} for a review).  For example,  gelatinous beads are seeded with yeast cells and used in reactors to produce ethanol \cite{yeastmatrix}.  To better understand the growth dynamics and physical properties of these systems,  it is important to characterize the nutrient transport properties of cell clusters as a function of both single cell nutrient uptake kinetics and the geometry of specific cell packings.

A nutrient concentration $\psi$ in some medium, such as water or gel, with a constant diffusion coefficient $D_0$  obeys the diffusion equation
\begin{equation}
\partial_t \psi = D_0 \nabla^2 \psi. \label{diffeq}
\end{equation}
In the steady state, the left hand side of Eq.~\ref{diffeq}  vanishes, and the equation reduces to Laplace's equation.  This is an important and well-studied equation in electrostatics, as it is the equation for the electrostatic potential $\Phi(\mathbf{r})\equiv\psi(\mathbf{r})$ in regions of space without any charges.  

In the case of nutrient diffusion, Eq.~\ref{diffeq} must be satisfied everywhere in the medium outside of the cell.  However, to completely solve Eq.~\ref{diffeq} in the steady state, we must specify boundary conditions.  One natural boundary condition is to set the concentration field at infinity to some constant value, i.e., $\psi(|\mathbf{r}| \rightarrow \infty) \rightarrow \psi_{\infty}$,  corresponding to a large nutrient bath with a uniform concentration $\psi_{\infty}$.  We also need boundary conditions on each cell surface.  For example, if the cell is a perfect nutrient absorber, then for all points $\mathbf{r}$ on the cell surface $S$, the nutrient concentration vanishes, i.e. $\psi(\mathbf{r}) = 0$.  In the electrostatic analogy, this condition  means zero electrostatic potential on every cell surface, i.e.,  each cell is a perfect grounded conductor.  Conversely, if the cell does not absorb any nutrient, (i.e. it is a perfect reflector), then Fick's first law of diffusion tells us that the derivative of the concentration along a direction $\hat{\mathbf{n}}$ perpendicular to the cell surface must vanish.  More precisely, the local nutrient flux density $J(\mathbf{r})$  into the cell at some point $\mathbf{r} \in S$ on the surface satisfies
\begin{equation}
J(\mathbf{r}) =   D_0 \hat{\mathbf{n}} \cdot \nabla \psi (\mathbf{r}) \label{fluxdef}
\end{equation}
so that $J(\mathbf{r}) = 0$ for all $\mathbf{r} \in S$ implies $ \left. \hat{\mathbf{n}} \cdot \nabla \psi (\mathbf{r})\right|_{\mathbf{r} \in S} = 0$.  In the electrostatic analogy, this would correspond to a perfect insulator with no surface charge, with a vanishing normal electric field. Of course, living cells are neither perfect absorbers nor perfect reflectors. A more realistic boundary condition  interpolates between these two ideal cases.  

A  boundary condition on the cell can be derived from a more microscopic model of the nutrient transporters.  For example, Berg and Purcell modeled transporters as small perfectly absorbing disks on the surface of an otherwise reflecting cell \cite{bergBPJ,bergBook}.  They showed that the cell requires very few transporters to act as an effectively perfect absorber: A cell with as little as a $10^{-4}$  fraction of its surface covered by transporters takes in half the nutrient flux of a perfect absorber!  Zwanzig and Szabo later extended this result to include the effects of transporter interactions and partially absorbing transporters \cite{zwanzig, zwanzig2}.  They showed that a homogeneous and partially absorbing cell surface model captures the average effect of all the transporters.  As discussed below, in many cases of biological interest, the cell cannot be treated as a perfect absorber.
The same partially absorbing boundary condition used by Zwanzig and Szabo will be derived in a different way in the next section.

Although Eq.~\ref{diffeq} is easily solved in the steady state for a single, spherical cell with the appropriate boundary conditions \cite{bergBPJ,bergBook}, the complicated arrangement of cells in a typical multi-cellular system, such as a yeast cell colony, implies a complex boundary condition that makes an exact solution  intractable -- one would have to constrain $\psi(\mathbf{r})$ and its normal derivative on a highly irregular object, like the surface of a cluster of grapes.  In this paper, we explore an ``effective medium'' approximation  to the exact solution of this problem.    

Effective medium theory treats a cluster of cells, or nutrient sinks, as a region with uniform effective nutrient transport properties (such as an effective diffusivity and nutrient absorption constant) that depend on the arrangement of cells in the cluster and the individual cell nutrient absorption properties.  A key feature of the effective medium theory is that these effective transport properties are derived in a self-consistent way.  These theories have been used to calculate many effective properties in heterogeneous systems such as conductivity, elasticity, and reaction rates (see \cite{cukierR, calefR} and Chapter 18 in \cite{torquatoBook} for reviews).     

The paper is organized as follows: We develop our theoretical model for nutrient uptake in single cells, dilute cell clusters, and dense clusters in Sec.~\ref{STheory}.  In Sec.~\ref{SSimulations} we compare our analytic results with numerical solutions of Eq.~\ref{diffeq} in the steady-state for clusters with hundreds of partially absorbing cells.  We discuss experimental tests of our model in Sec.~\ref{SExperiments} and provide concluding remarks in Sec.~\ref{SConclusions}.

\section{\label{STheory} Theoretical Model}

We  now discuss how to couple  single cell nutrient uptake kinetics to the nutrient uptake behavior of an entire cluster or colony of cells via effective medium theory. In what follows we assume the cells are all identical and spherical.  Although our experimental model is the budding yeast cell, the theoretical treatment is quite general and can be adapted to any cell cluster that absorbs nutrients that reach it via diffusion.

\subsection{\label{SingleCell}Single Cell Nutrient Uptake}

\begin{figure}
\includegraphics[height=1.9in]{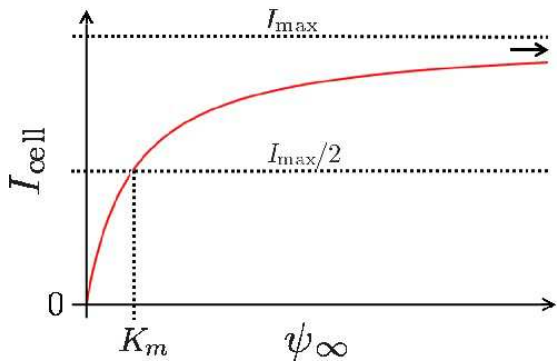}
\caption{\label{MMfig} (Color online) A plot of the nutrient current $I_{\mathrm{cell}}$ (solid red line) into a cell as a function of the ambient nutrient concentration $\psi_{\infty}$, which follows Michaelis-Menten kinetics.  The maximum current $I_{\mathrm{max}}$ and the kinetics parameter $K_m$ are also shown.}
\end{figure}

We first review nutrient uptake by a single cell, with nutrient transporters following Michaelis-Menten kinetics.  We smooth out the effect of a discrete set of transporters within the cell wall and study a radially symmetric model of nutrient uptake. A Michaelis-Menten model then  means that the total nutrient current $I_{\mathrm{cell}}$ into a cell is related to the ambient nutrient concentration $\psi_{\infty}$ via
\begin{equation}
 I_{\mathrm{cell}}  = \frac{I_{\mathrm{max}} \psi_{\infty}}{K_m+\psi_{\infty}}, \label{MichaelisMenten}
\end{equation}
where $I_{\mathrm{max}}$ (sometimes called $V_{\mathrm{max}}$ in the literature) is the saturating nutrient current into the cell as $\psi_{\infty} \rightarrow \infty$ and $K_m$ is the concentration at which $I_{\mathrm{cell}} = I_{\mathrm{max}}/2$, as shown in Fig.~\ref{MMfig}.  Even if  a cell does not obey these kinetics for all $\psi_{\infty}$, one can often define a range of concentrations $\psi_{\infty}$ characterized by effective kinetic parameters $I_{\mathrm{max}}$ and $K_m$,  a characterization often used  in experimental studies of nutrient uptake. Many studies infer an effective $I_{\mathrm{max}}$ and $K_m$ from the measured amount of nutrient consumed by a dilute suspension of cells \cite{meijer, srienc, gramp}.

The parameters $I_{\mathrm{max}}$ and $K_m$ are determined by the microscopic kinetics of each individual cell transporter and the density of these transporters on the cell surface.  A third contribution arises from the structure of the cell wall, which influences the rate of transport of the nutrient into the cell.  We can find $I_{\mathrm{max}}$ and $K_m$ by appealing to a microscopic model of the transporters.       Berg and Purcell, for example,  modeled each transporter as a small, perfectly absorbing disk on the surface of a perfectly reflecting cell \cite{bergBook,bergBPJ}.    Zwanzig showed that the large scale physics of this model are approximated by a partially absorbing boundary condition on the surface of the cell \cite{zwanzig}. We will now derive this boundary condition in another way and connect $I_{\mathrm{max}}$ and $K_m$ to our model.  
\begin{figure}
\includegraphics[height=1.9in]{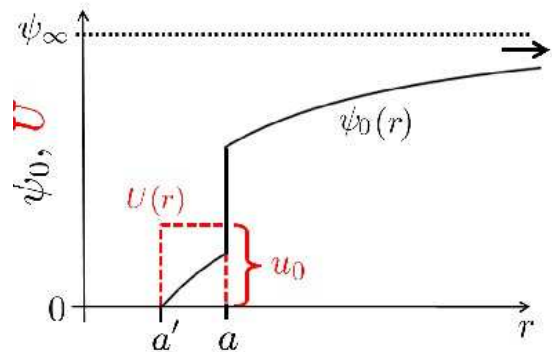}
\caption{\label{potentialbarrierfig} (Color online) A radially symmetric model of nutrient uptake in a single cell. We specify a steady concentration $\psi_{\infty}$ far away from the cell. The cell is centered at the origin and has radius $a$.  The steady state concentration profile $\psi_0(r)$ (black solid line) is calculated for a rectangular potential barrier  (red dashed line) of height $u_0$  (designed to model the complex uptake dynamics of the cell wall) and  a perfectly absorbing nutrient sink at $r = a'$. The concentration profile approaches $\psi_ {\infty}$ at large $r$ and exhibits a jump discontinuity at $r=a$.  The nutrient currents are constrained to be continuous.}
\end{figure}

To model the cumulative effect of many microscopic details, such as transporter kinetics, the cell wall, etc., on the nutrient uptake we introduce a  potential energy barrier $U(\mathbf{r})$.  Once a nutrient molecule surmounts this barrier, it gets absorbed by the cell.    The nutrient concentration $\psi(\mathbf{r})$  obeys the Fokker-Planck equation
\begin{equation}
\partial_t \psi =D_0 \nabla \cdot  \left[  \, \nabla \psi + \frac{1}{k_B T} \, (\nabla U) \, \psi \right], \label{FPE}
\end{equation}
where $k_B$ is the Boltzmann constant and $T$ is the temperature of the nutrient solution \cite{vankampen}.  Simple diffusion is recovered when the potential is constant. For simplicity, let's suppose that the nutrient must overcome a radially symmetric potential barrier $U(r)$ that has a rectangular ``lip'' of height $u_0$ at $r = a$  and with width $w \equiv a-a'$ (see Fig.~\ref{potentialbarrierfig}).  Thus,
\begin{equation}
U(r) = \begin{cases} 0 & r > a \\ u_0 & a' \leq r \leq  a \\
0 & r< a' \end{cases},
\end{equation}
where  $u_0$ is the height of the barrier.     We assume perfect absorption at $a'$ and a constant nutrient concentration infinitely far away,
\begin{align}
 \begin{cases} \psi(r=a',t) = 0 \\
 \psi(r \rightarrow \infty, t) = \psi_{\infty} .
 \end{cases} \label{singlecellBCs}
 \end{align}

 To determine the nutrient flux into the cell, we solve Eq.~\ref{FPE} for the steady state profile $\psi_0(r)$.  In addition to the boundary conditions (Eq.~\ref{singlecellBCs}), we ensure the continuity of the nutrient flux at $r = a$ via the ``jump conditions'' at $r = a$, as discussed in Ref. \cite{risken}.   The resulting concentration profile  reads
\begin{align}
\psi_0(r) = \begin{cases}
\psi_{\infty}-\dfrac{a a' \psi_{\infty}}{a'+(a-a') e^{ u_0/k_B T}} \,\dfrac{1}{r} & r \geq a \\[15pt]
\dfrac{a (r-a')  \psi_{\infty}}{\left(a'+(a-a') e^{ u_0/k_B T}\right) } \, \dfrac{1}{r} & a' \leq r < a,
\end{cases} \label{SSFPEsolution}
\end{align}
 with $\psi_0(r) = 0$ for $r < a'$.  The shape of the solution $\psi_0(r)$ is shown in Fig.~\ref{potentialbarrierfig}.
 
  Let's now consider narrow potential barriers relative to the cell radius ($w = |a-a'| \ll a$).  Then, from Eq.~\ref{SSFPEsolution}, we find the concentration gradient just outside the cell surface: 
 \begin{align}
 \left. \partial_r \psi_0 \right|_{r \rightarrow a^+} \approx \frac{1
}{w} \exp \left[-\frac{u_0}{k_B T} \right]    \psi_0 (a^+)\equiv \kappa  \psi_0 (a^+), \label{radiationBC1}
 \end{align}
 where the $+$ superscript indicates that we take the limit $r \rightarrow a$ from outside the cell.   
Eq.~\ref{radiationBC1} reveals that the  gradient of $\psi_0$ normal to the cell surface is proportional to  $\psi_0(r)$ just outside.  Notice that $\kappa \rightarrow \infty$ when $w \rightarrow 0$ (we also let $u_0 \rightarrow 0$), so that $\psi_0(a^+) \rightarrow 0$ at the cell surface to keep the flux finite.  Thus, the cell is perfectly absorbing within our model if there is no potential barrier.  Similarly, for a very large barrier $(u_0 \rightarrow \infty,w \mbox{ finite})$, we have $\kappa \rightarrow 0$ so that there is no flux of nutrient into the cell and $\partial_r \psi_0(a^+) \rightarrow 0$, signifying a perfect reflector.  

The proportionality between a field and its gradient at a boundary  is called a \textit{radiation} boundary condition in the physics literature and can be derived quite generally \cite{menon}.  This boundary condition is a natural coarse-grained description of the Berg and Purcell model of transporters as absorbing disks.  Zwanzig and Szabo \cite{zwanzig,zwanzig2}  have used the radiation boundary condition to successfully model the physics of both perfectly and partially absorbing disks on scales larger than the disk spacing, thus confirming our expectation that   the coarse-grained nutrient uptake can be modeled by the ubiquitous radiation boundary condition with an appropriate choice of $\kappa$.

The absorptive strength of the cell can be parameterized by the dimensionless number $\nu \equiv \kappa a$, where $a$ is the cell radius.  In chemical engineering, $\nu$ is sometimes referred to as a Sherwood number \cite{sherwood}.  If $P_{\mathrm{abs}}$ is the probability that a nutrient particle at the cell surface will be absorbed by the cell (instead of escaping to infinity), then first-passage techniques from probability theory \cite{redner} lead to
\begin{equation}
P_{\mathrm{abs}} = \frac{\nu}{1+\nu}.
\end{equation}
Thus, $\nu = \kappa a \ll 1$ indicates poor nutrient absorption while  $\nu \gg 1$ indicates a good absorber.  Note that at $\nu = 1$, the nutrient has equal probability of being absorbed at the cell surface or escaping to infinity. 

We now connect $\nu$ with the measurable biological parameters $I_{\mathrm{max}}$ and $K_m$.  Recall that the nutrient flux into a single cell is related to the ambient nutrient concentration $\psi_{\infty}$  via the Michaelis-Menten relation Eq.~\ref{MichaelisMenten}.  Suppose for now that the cells are well-separated, so the nutrient uptake $I_{\mathrm{cell}}$ of any given cell is independent of the others (i.e., there is no nutrient shielding), and that the ambient nutrient concentration $\psi_{\infty}$ is held constant, so that the parameters $I_{\mathrm{max}}$ and $K_m$ at each cell can assume their steady-state values.  We  assume the nutrient solution experiences no macroscopic flows, such as convection currents, that would bias the isotropic absorption kinetics of the cell.  The nutrient concentration $\psi(r \equiv |\mathbf{r}|)$ then satisfies $\nabla^2 \psi(\mathbf{r}) = 0$ in the steady state, with the boundary conditions $\psi(r \rightarrow \infty) = \psi_{\infty}$ and $ \left. \hat{\mathbf{n}} \cdot \nabla \psi \right|_S= \left. \kappa \psi \right|_S$ at each cell surface $S$, as discussed above.  

Upon inserting Eqs.~\ref{SSFPEsolution} and \ref{radiationBC1} into  Fick's first law \cite{redner} (see also Eq.~\ref{fluxdef}), we find the steady-state nutrient current into an individual cell,
\begin{equation}
I_{\mathrm{cell}} = \int_S D_0 \hat{\mathbf{n}} \cdot \nabla \psi(r) a^2 \, \mathrm{d} \Omega \approx  \frac{4 \pi D_0 \psi_{\infty}  \nu a}{1+\nu}, \label{cellflux}
\end{equation}
 where we integrate over the surface of the cell $S$ (so that $\mathrm{d}\Omega = \sin \theta \, \mathrm{d} \theta \, \mathrm{d}\phi$ in spherical coordinates) and again assume $w = a-a' \ll a$. Comparison of Eq.~\ref{cellflux} with Eq.~\ref{MichaelisMenten} leads to $\nu$ as a function of biological parameters.  In the limit of low ambient nutrient concentration $(\psi_{\infty} \ll K_m)$, we have
\begin{align}
\nu = \frac{I_{\mathrm{max} }}{4\pi a   D_0 K_m-I_{\mathrm{max} }}.
\end{align}
It is also possible to define an effective $\nu$ for a reflecting spherical cell uniformly covered by identical, partially absorbing disks with radius $a_{\mathrm{disk}}$ and absorption parameter $\kappa_{\mathrm{disk}}$.   In this case, using the boundary condition Eq.~\ref{radiationBC1} on each disk surface,  Zwanzig and Szabo find that the effective parameter $\nu$ for the entire cell (for $a_{\mathrm{disk}}\kappa_{\mathrm{disk}} \ll 4/\pi$) is $\nu = N_{\mathrm{disk}} a_{\mathrm{disk}}^2  \kappa_{\mathrm{disk}}/4 a$, where $N_{\mathrm{disk}}$ is the number of disks on the cell surface (see Ref.~\cite{zwanzig2}).  

 A rough estimate of $\nu$ for glucose uptake by a \textit{S. cerevisiae} cell follows from values for $I_{\mathrm{max}}$ ($4.2 \times 10^7$~molecules/sec), $K_m$ ($7.4$ mM), $a$ (2~$\mu$m), and $D_0$ (670 $\mu$m$^2$/sec) found in the literature \cite{meijer, yeastdimensions, glucosediff}.  The $I_{\mathrm{max}}$ is particulary difficult to estimate as the nutrient uptake rate in experiment is calculated per gram of dried yeast taken out of a liquid culture.  To get an uptake rate per cell, we estimate that a yeast cell has a 2 pg dry weight \cite{dryweight}. We find that these yeast cells are in fact very poor absorbers with $P_{\mathrm{abs, \, yeast}} \approx \nu_{\mathrm{yeast}} \approx 6 \times 10^{-4} \sim 0.001$ within an order of magnitude.  Gram-negative bacterial cells differ substantially. Again using literature values for $I_{\mathrm{max}}$ ($2 \times 10^7$~molecules/sec),  $K_m$ (1 $\mu$M), and  $a$ (0.5 $\mu$m) \cite{srienc, ecoliradius}   for a single \textit{Escherichia coli} cell, we find that $P_{\mathrm{abs, \, gram-n.}} \approx \nu_{\mathrm{gram-n.}}  \approx 0.09\sim 0.1$.  Thus, this bacterium is $\sim 100$ times more absorbent than the yeast cell:  This striking difference has profound biological implications since, as we will show in the next sections, the parameter $\nu$ greatly influences the growth of a cell colony. 

 The large disparity in $\nu$ values may be due to the thicker cell walls of \textit{S. cerevisiae}, compared to gram-negative bacteria like \textit{E. coli}.       The presence of a cell wall can have two effects. First, the diffusion coefficient of the transported nutrients may be lower in the cell wall medium  than it is in the bulk solution.  Second, the absorbing surface lies at the plasma membrane, not the surface of the wall. The second effect implies that the absorbing surface is at a distance $a-w$ from the cell center (rather than $a$), where $a$ is the cell radius and $w$ the thickness of the wall. So, even if the diffusion coefficient inside the wall is the same as in the bulk and we have a perfectly absorbing plasma membrane, the effective value of $\nu$ is  $\nu = a/w - 1$. This argument is consistent with the measured glucose uptake kinetics for the gram-positive bacterium \textit{Luconostoc mesenteroides}, which has  a   cell wall  thicker than \textit{E. coli} and thinner than \textit{S. cerevisiae}.  We find $P_{\mathrm{abs, \, gram \, p. }} \approx \nu_{\mathrm{gram \, p.}} \approx 0.05$  from the literature values of the parameters \cite{gramp,grampsize}.   Of course, other factors apart from the cell wall thickness could be relevant.

The experimental results discussed here for single cells tell us that it is important to consider a large range of the parameter $\nu = \kappa a$:   $\nu$ can range over at least two orders of magnitude $(0.001< \nu < 0.1)$ for  yeast, gram-positive, and gram-negative bacteria.  Thus, neither  perfectly reflecting nor perfectly absorbing boundary conditions are relevant for nutrient uptake in many cell populations.  Instead, we develop a theory for the nutrient absorption by a  cluster of cells with arbitrary $\nu$.  This is the subject of the next section.

\subsection{\label{CellClusterUptake}Nutrient Uptake in Cell Clusters}

We now consider nutrient absorption by a cluster or colony of cells.  In general, this is a very complicated problem involving solving the diffusion equation in the interstitial area of the cluster while making sure the  boundary condition (Eq.~\ref{radiationBC1}) is satisfied at each cell surface.  Although analytical results are possible for a single cell, we must resort to numerical solutions and approximations when dealing with a cluster.  A typical 1~mm diameter yeast cell colony contains over $10^6$ cells, making the exact solution for nutrient uptake by such a colony intractable even numerically.

\begin{figure}
\includegraphics[width=3.2in]{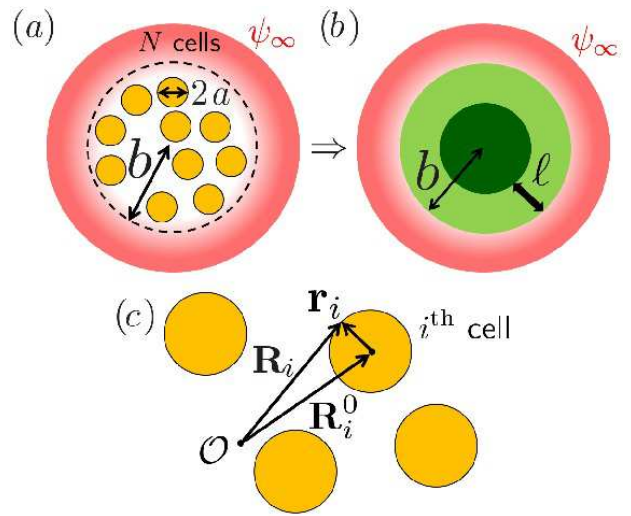}
\caption{\label{feffectivemedium} (Color online) Schematic of the effective medium approximation, which replaces $N$ orange (gray) cells in a spherical cluster with radius $b$ (enclosed by the dashed lines) in $(a)$ by a homogeneous, attenuating medium in $(b)$ shown in dark and light shades of green. The lightly shaded green rim of width $\ell$ illustrates the section of the cluster receiving enough nutrients to grow. $\psi_{\infty}$ is the limiting nutrient concentration in the variably shaded red region, outside the cluster. To  analyze this problem, we use the coordinate system in $(c)$, with an origin at $\mathcal{O}$, cell centers located at $\{\mathbf{R}^0_i\}_{i=1}^N$, and vectors $\mathbf{r}_i$ pointing from the cell center to the surface of the cell. }
\end{figure}
To model nutrient uptake (and nutrient shielding) in a group of cells, we again consider the steady-state diffusion equation with boundary conditions provided by a disorderly cluster of $N$ identical spherical cells all with radius $a$ and no overlaps (see Fig.~\ref{feffectivemedium}$(a)$).  Let the cells be located  inside a spherical region of radius $b \gg a$ with centers at positions $\mathbf{R}_i^0$, where $i=1,\ldots,N$.  We also employ local spherical coordinates at each cell to write vectors $\mathbf{r}_i=(a,\Omega_i)$ pointing at positions on the cell surface, where  $\Omega_i \equiv (\theta_i,\phi_i)$  is a pair of polar angles that specifies the direction of $\mathbf{r}_i$ relative to the center of the $i^{\mathrm{th}}$ cell (see Fig.~\ref{feffectivemedium} $(c)$). Consider a particular configuration of these $N$ cells and denote by $\Psi(\mathbf{r})$ the \textit{exact} solution to the steady-state concentration field in the interstitial region for this particular configuration.

   Upon assuming a time-independent steady-state, we modify Eq.~\ref{diffeq} to include a set of Lagrange multiplier functions $\sigma_i(\mathbf{r})$ defined on the cell walls,
\begin{align}
D_0 \nabla^2 \Psi(\mathbf{r}) = \sum_{i=1}^N \int  \sigma_i(\mathbf{r}_i) \, \delta(\mathbf{r}-\mathbf{R}_i) \, \mathrm{d} \Omega_i +s(\mathbf{r}). \label{exactdiffeq}
\end{align}
 We also set $\mathbf{R}_i \equiv \mathbf{R}_i^0+\mathbf{r}_i$ (see Fig.~\ref{feffectivemedium}$(c)$) and choose $\mathbf{r}$ to be inside the cell cluster.  The Lagrange multiplier functions $\{\sigma_i(\mathbf{r}_i)\}$ will be chosen to satisfy the radiation boundary condition at each cell surface (the functions $\{ \sigma_i (\mathbf{r}_i) \}$ would be charge densities in electrostatics).  These boundary conditions (given  Eq.~\ref{radiationBC1} with $\nu = \kappa a$ for a single cell) read, for all $i$ and $\mathbf{R}_i$,
\begin{align}
\Psi(\mathbf{R}_i) = \nu^{-1} \, \mathbf{r}_i \cdot \nabla \Psi(\mathbf{R}_i) \equiv \mathcal{Q}_i \Psi(\mathbf{R}_i), \label{radiationBC2}
\end{align}
 where $\mathcal{Q}_i \equiv \nu^{-1}\, \mathbf{r}_i \cdot \nabla$ is a convenient gradient operator used in the detailed effective medium calculation in Appendix~\ref{AEM}. The  source function $s(\mathbf{r})$  allows us to incorporate additional boundary conditions on the concentration field. 

We now average over all possible cell configurations (consistent with excluded volume interactions between cells) to obtain the average transport properties of the nutrient in a cell cluster.  Upon averaging both sides of Eq.~\ref{exactdiffeq}, we find
\begin{align}
D_0 \nabla^2 \psi(\mathbf{r}) & =    \left\langle \sum_{i=1}^N \int_{S_i} \sigma_i(\mathbf{r}_i) \, \delta(\mathbf{r}-\mathbf{R}_i) \,\mathrm{d} \Omega_i   \right\rangle+s(\mathbf{r}) \label{clusterdiffeq}\\
 & \approx   \int  \Sigma(\mathbf{r}'-\mathbf{r}) \, \psi(\mathbf{r}') \, \mathrm{d} \mathbf{r}' +s(\mathbf{r}), \label{linresponse}
\end{align}
where $\psi(\mathbf{r}) \equiv \left\langle \Psi (\mathbf{r}) \right\rangle$. The bracket average is an ensemble average over cell configurations and  $\Sigma(\mathbf{r})$ is a linear response function describes how the cells deform the concentration field.  

The linear response approximation, justified here by comparisons with simulations, is only valid  away from the cluster edges and for sufficiently small concentration field deformations.  In addition, this approximation often breaks down for time-dependent diffusion \cite{richards} because the transient diffusive dynamics are dominated by slowly decaying modes due to large voids inside the cell cluster \cite{kayser}.  Ref.~\cite{cukier2} uses a more microscopic description of nutrient diffusion to examine the validity of the linear response approximation in more detail.  We will forgo these complications here and exploit the linear approximation above, checking our assumptions using experiments and simulations, as had been done for the physics of fluorescence quenching \cite{kruger}.

An exact evaluation of $\Sigma(\mathbf{r})$ involves an ensemble average denoted by brackets in Eq.~\ref{clusterdiffeq}. This average requires the full probability distribution $P( \{ \mathbf{R}_i^0 \} _{i=1}^N )$ of observing $N$ cells with centers $\{ \mathbf{R}_i^0 \}_{i=1}^N$.  Unfortunately, an exact solution obtained in this way would require knowledge of all of the correlations between the cell positions, which may not be experimentally accessible. We will assume for now that this distribution is known.  Later, the effective medium theory developed in Sec.~\ref{SSEffectiveMedium} will approximate $\Sigma(\mathbf{r})$ in a self-consistent way using just the one and two cell center distributions.

What happens to the configurationally averaged solution $\psi(\mathbf{r})$ of Eq.~\ref{linresponse} over distances large compared to the size of a single cell?  Specifically,  how does the cell colony absorb nutrients {\it on average}, as if it were the homogeneous medium illustrated in Fig.~\ref{feffectivemedium}$(b)$?  To answer this question, we perform a gradient expansion of Eq.~\ref{linresponse}. As discussed in Ref. \cite{cukier2}, such a gradient expansion neglects intrinsically non-local contributions  to $\Sigma(\mathbf{r})$ due to the fluctuations in the concentration field that cannot be averaged over large distances.  The mean field approximation used here then uses the resulting transport coefficients to describe the absorptive properties of the cluster.  Simulations have shown that this approach correctly models the physics in  related systems \cite{forney}.     

  Eq.~\ref{linresponse} in Fourier space reads
\begin{equation}
- D_0 q^2 \psi(\mathbf{q}) = \Sigma(\mathbf{q}) \psi(\mathbf{q})+s(\mathbf{q}) , \label{effectivepropagator}
\end{equation}
where we have applied the convolution theorem and all the functions are now their Fourier transformed functions of the 3d wave-vector $\mathbf{q}$.  On average, the cells in the cluster should be distributed isotropically and  $\Sigma(\mathbf{q})$ can only depend on $q \equiv |\mathbf{q}|$.  Expanding $\Sigma$ around $q = 0$ gives us the desired gradient expansion:
\begin{align}
- D_0 q^2 \psi(\mathbf{q}) & = \Sigma(q=0) \psi(\mathbf{q})+ \frac{\Sigma''(q=0)q^2\psi(\mathbf{q})}{2} \nonumber \\
& \qquad \qquad +\mathcal{O}(q^4)+s(\mathbf{q}) \nonumber \\
& \equiv k \psi(\mathbf{q}) + \delta D q^2 \psi(\mathbf{q})+\mathcal{O}(q^4)+s(\mathbf{q}), \label{qexpansion}
\end{align}
where we have identified an absorptive term $k \psi$ and a correction to the diffusion term $\delta D q^2 \psi$.  We neglect higher powers of $q$ (i.e., higher order derivatives of the concentration field) in our coarse-grained reaction-diffusion description of nutrient transport in a large colony of cells.  Upon returning to real space, we find the desired macroscopic transport equation for the configurationally averaged nutrient concentration:
\begin{equation}
 D \nabla^2 \psi (\mathbf{r})- k \psi (\mathbf{r})= s(\mathbf{r}) \label{rxndiffeq}
\end{equation}
where $D = D_0 + \delta D$.  The coefficients $D$ and $k$ characterize the macroscopic diffusion and absorption, respectively, of the nutrient in the cell colony. The crucial question is how the effective diffusion constant $D$ and absorption $k$ depend on parameters such as $\nu$ and the cell volume fraction $\phi$.

In general, we expect that $k$ will be a positive, increasing function of the cell volume fraction $\phi$, since the cluster becomes more absorbent as we introduce more cells. It will also increase with $\nu = \kappa a$ as each cell will absorb more nutrients as $\nu$ increases.  The sign of $\delta D$ is more subtle, because of two competing factors: The confinement of the nutrient by the cells in the interstitial space will \textit{decrease} the effective diffusion constant $D$, while the nutrient gradients in random directions induced by nutrient uptake by cells will \textit{increase}  $D$.  Since $\nu$  controls the amount of absorption, we expect that (for small $\phi$) $\delta D < 0$ for $\nu \ll 1$ and $\delta D > 0$ for $\nu \gg 1$.  This is confirmed by our effective medium calculation of $D$ in Sec.~\ref{SSEffectiveMedium}.

We assumed  that the radii of the cells in the cluster are monodisperse.  This approximation might not be realistic in cell clusters at different stages of their cell cycle.  However, in both polydisperse and monodisperse cases, we can still define a cell volume fraction $\phi$.  The effective transport coefficients $D$ and $k$ will depend on this volume fraction and the polydispersivity of the radii.  For a fixed $\phi$, polydisperse cell radii will decrease the total cell surface area and, consequently,  \textit{decrease} the nutrient absorption $k$.  For example, the ratio of the total cell surface  area $S_{\mathrm{poly}}$ of a  cluster of cells with a Gaussian  radius distribution (with average $\langle a \rangle$ and variance $\sigma_a \ll \langle a \rangle$) to the total surface area $S_{\mathrm{mono}}$ for a  cluster of cells  with the same radius $\langle a \rangle$ is given by \cite{torquatoBook}
\begin{equation}
\frac{S_{\mathrm{poly}}}{S_{\mathrm{mono}}} = \frac{ \left\langle a^2 \right\rangle \left\langle a \right\rangle}{ \left\langle a^3 \right\rangle} =  \frac{\sigma^2_a+ \left\langle a \right\rangle^2}{3 \sigma^2_a+ \left\langle a \right\rangle^2} < 1.
\end{equation}
Thus, a theory of monodisperse cells is reasonable provided $\langle a \rangle \gg  \sigma_a$.  Also, we don't expect the correction to vary much with $\nu$ since the main effect of polydispersivity seems to be from the surface area decrease.  Finally, computer simulations for two discrete sphere sizes reveal that the ratio $k_{\mathrm{bidisp}}/k_{\mathrm{mono}}$ of absorption coefficients for   (perfectly absorbing) cells satisfies $k_{\mathrm{bidisp}}/k_{\mathrm{mono}} \approx (S_{\mathrm{bidisp}}/S_{\mathrm{mono}})^2$    \cite{torquatoPoly, zhengPoly} over a wide range of packing fractions.  Thus, it may be possible to   approximate polydispersity in our theory by  reducing the absorption coefficient $k$ by such a geometric factor. However, unless stated otherwise, we henceforth  ignore this complication and instead consider cell clusters with monodisperse radii.

\subsection{\label{PenetrationDepth} The Macroscopic Screening Length $\xi$}

 An important nutrient screening length associated with Eq.~\ref{rxndiffeq} for cell clusters is
\begin{equation}
\xi \equiv \sqrt{\frac{D}{k}}. \label{xilength}
\end{equation}
We now use effective medium theory to calculate $\xi$ and relate this length to cell configurations and single cell nutrient uptake kinetics.

Consider first a spherical cell cluster (or ``colony'') of radius $b$ in which  $\psi$ satisfies the effective medium result Eq.~\ref{rxndiffeq}.  Assume as well that $ \nabla^2 \psi =0$ outside of this spherical region.  Then, with the boundary condition $\psi(r \rightarrow \infty) = \psi_{\infty}$ and continuity of the nutrient current at $r = b$, we find  
\begin{equation}
\psi(r)  = \begin{cases} \dfrac{D_0\psi_{\infty} \xi}{Dr\cosh(b/\xi)} \, \dfrac{\sinh (r/\xi)}{1+  (D_0/D-1) T(b/\xi)}  & r \leq b \\[15pt]
\psi_{\infty} \left[1+ \dfrac{  b }{r} \left[ \dfrac{T(b/\xi)-1}{1+(D_0/D-1)T(b/\xi)  } \right] \right] &r > b
\end{cases}, \label{psisol}
\end{equation}
where $T(x) \equiv  \tanh(x)/x$.  When $b \gg \xi$, the concentration field $\psi(r) \approx \frac{D_0 \psi_{\infty} \xi }{D r} \exp[(r-b)/\xi]$ near the colony surface, and thus decays exponentially as we move further into the interior.  Thus, $\xi$ is characterstic $e$-folding length of the nutrient decay.

  Eq.~\ref{psisol} leads to the total flux $I_{\mathrm{cluster}}$ into the cell \textit{cluster} in the same way as the single cell discussed above.  We will see in the next two sections that a good approximation is $D \approx D_0$.  Then, the nutrient flux into the cluster is
 \begin{equation}
 I_{\mathrm{cluster}} = 4 \pi D_0 \psi_{\infty} \left[ b- \xi \tanh \left( \frac{b}{\xi} \right) \right]. \label{clusterflux}
 \end{equation}
 Thus, when $\xi \ll b$ so the nutrient does not penetrate far into the cell colony,  $I_{\mathrm{cluster}} \approx 4 \pi D_0 \psi_{\infty} b$ and the entire colony acts as if it is a single, perfectly absorbing sphere with radius $b$. Conversely, if $\xi \gg b$, then nutrient penetrates deep into the colony.  In this limit we have
\begin{equation}
I_{\mathrm{cluster}} = \frac{4 \pi D_0 \psi_{\infty} b^3}{3 \xi^2} + \mathcal{O}(\delta^4) , \qquad \delta = \frac{b}{\xi} \ll 1, \label{fluxlargexi}
\end{equation}
so that the total nutrient absorption is down relative to a perfectly absorbing colony by a factor of $\delta^2/3 \ll 1$.   $I_{\mathrm{cluster}}$ now scales with colony volume $(\propto b^3)$, since each  cell in the colony contributes to the nutrient uptake.

\begin{figure}
\includegraphics[height=2.2in]{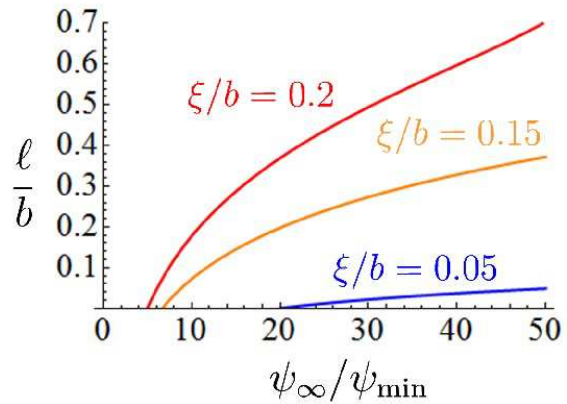}
\caption{\label{thicknesslplot}(Color online) Plot of the thickness $\ell$ of actively growing cells in a spherical cluster of size $b$ as a function of the nutrient bath concentration $\psi_{\infty}$ for various values of the penetration depth $\xi$.}
\end{figure}

A related biologically relevant parameter is the thickness $\ell$ of the outer shell of actively growing cells, as illustrated in Fig.~\ref{fconfocal} and by the lightly shaded green band in Fig.~\ref{feffectivemedium}$(b)$:  Suppose that cells require some minimum concentration $\psi_{\mathrm{min}}$ of nutrient in order to grow (or exhibit some level of growth-coupled fluorescence).  We can estimate $\ell$  by finding the location $r_{\mathrm{min}}$ inside a spherical colony such that  $\psi(r_{\mathrm{min}}) = \psi_{\mathrm{min}}$ (see Eq.~\ref{psisol}) and setting $\ell \equiv b- r_{\mathrm{min}}$.  For $b \gg \xi$ and $D \approx D_0$, this length is related to $b$, $\xi$, $\psi_{\mathrm{min}}$ and $\psi_{\infty}$ via (see Fig.~\ref{thicknesslplot})
\begin{equation}
\ell \approx b+ \xi W_{-1} \left[ - \frac{\psi_{\infty}}{\psi_{\mathrm{min}}} \, e^{-b/\xi} \right], \label{ellatlargeb}
\end{equation}
where $W_{-1}(x)$ is one of the branches of the Lambert-$W$ function \cite{LambertW}.   The thickness $\ell$ can also be determined experimentally using a fluorescence reporter, as shown in Fig.~\ref{fconfocal}.   Experimental estimates of $\ell$ in yeast cell colonies are discussed in Sec.~\ref{SExperiments}.

\subsection{\label{SSDiluteCells} The Dilute Cell Limit}

As mentioned in Sec.~\ref{SIntro}, dilute clusters of microbial cells can be realized in experiment by embedding the cells in a  matrix.   The initial cell density in such a matrix  can be tuned over a wide range and is often quite small.  Moreover, since the  matrix usually has a negligible influence on the diffusion of small molecules like glucose \cite{diffmatrix}, we now develop an effective medium theory for clusters with cell densities low enough to neglect interactions.
 This theory will serve as an important and instructive limiting case.

We approximate the nutrient flux into each cell with the single cell result Eq.~\ref{cellflux}, replacing $\psi_{\infty}$ with the local value $\psi(\mathbf{r})$, where $\mathbf{r}$ is the location of the cell.  This approximation does not take into account the finite size of our cells or the deformation of the concentration field $\psi(\mathbf{r})$ around each cell. Thus, we cannot use this dilute limit to find a correction to the diffusion constant $D$.  However, as we will see in the next section, we expect this to be a very small correction, especially for small packing fractions $\phi$.     

\begin{figure}
\includegraphics[height=2in]{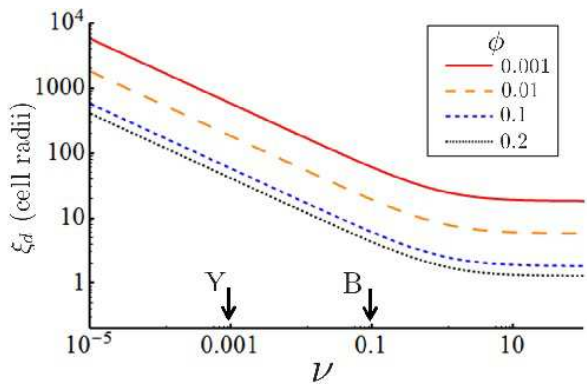}
\caption{\label{dilutexifig}(Color online) Plot of the screening length $\xi_d$ in the small $\phi$ limit as a function of $\nu$ for various values of $\phi$ on a log-log scale.  $\xi$ crosses over to the perfectly absorbing limit at $\nu \sim 1$.  The arrows Y and B denote, respectively, typical values of $\nu$ for baker's yeast and bacteria.}
\end{figure}

To approximate the nutrient absorption $k$,  we assume that each cell independently contributes to a nutrient flux per unit volume 
\begin{equation}
j(\mathbf{r}) =\frac{4 \pi D_0 \psi(\mathbf{r}) \nu an}{1+\nu}  = \frac{3 D_0 \phi \nu}{(1+\nu)a^2} \, \psi(\mathbf{r}),
\end{equation}
 where $n$ is the cell number density which we relate to the cell volume fraction $\phi = \frac{4 }{3} \,\pi  a^3 n$.    We conclude that the macroscopic transport coefficients in this dilute approximation are
\begin{equation}
D = D_0 \mbox{ and } k =  \frac{3 D_0 \phi \nu}{(1+\nu)a^2} . \label{diluteparameters}
\end{equation}

More sophisticated theories can find a correction to $D$ that is linear in $\phi$ and higher order corrections to $k$ (see Ref.~\cite{calefR} for a review).  Eq.~\ref{diluteparameters} implies that the screening length $\xi= \sqrt{D/k}$ in the dilute limit is given by
\begin{equation}
\xi \rightarrow \xi_d \equiv \sqrt{ \frac{1+\nu}{3 \phi \nu}} \,a. \label{dilutexi}
\end{equation}
Fig.~\ref{dilutexifig} shows $\xi_d$ as a function of the absorption parameter $\nu = \kappa a$: the screening length is a strong function of $\nu$ for $\nu < 1$ and crosses over to a perfectly absorbing regime when $\nu > 1$.  If the cells in a dilute cluster at low volume fraction have polydisperse radii in a Gaussian distribution of cell radii with average $\langle a \rangle$ and standard deviation $\sigma_a \ll \langle a \rangle$, Eq.~\ref{dilutexi} becomes
\begin{equation}
\xi_{d, \, \mathrm{poly}} = \sqrt{\frac{1+\nu}{3 \phi \nu}} \, \left[ \langle a \rangle^2 +3\, \sigma_a^2 \right]^{1/2}. 
\end{equation}
As expected, the reduced total cell surface area in the polydisperse cluster leads to a larger screening length.  Note that the ratio $\xi_{d,\,\mathrm{poly}}/\xi_d$ is independent of $\nu$ and  $\phi$ in this limit.  

For yeast cells with  $\phi \approx 0.1$ and $\nu \approx 0.001$, Eq.~\ref{dilutexi} predicts  $\xi \approx 60 a$.  Thus, glucose penetrates far into the yeast cell cluster when $\phi \approx 0.1$ (e.g., for yeast embedded in gel), allowing a substantial fraction of the yeast population to grow.  Conversely, we expect that $\xi \approx 6 a$ for a bacterial colony $(\nu \approx 0.1)$ at $\phi \approx 0.1$ volume fraction.  To treat cell arrangements like the one in Fig.~\ref{fconfocal}, we clearly need to go beyond the dilute limit and determine the dependence of $\xi$ on $\phi$ and $\nu$ more carefully:  Yeast and other cell colonies rarely grow at low volume fractions; cells typically clump together and pack themselves in an amorphous structure with a volume fraction approaching that of the random close packing density,  $\phi \approx 0.6-0.7$ \cite{RCP}.

\subsection{\label{SSEffectiveMedium}Effective Medium Theory for Dense Cell Clusters}

 For  $\phi \gtrsim\ 0.1$, the dilute approximation breaks down,  and we must solve Eq.~\ref{exactdiffeq} for  $\Psi(\mathbf{r})$ more exactly.   Upon defining the diffusive Green's function $G_0(\mathbf{r})  = (4 \pi D_0 r)^{-1}$, we can rewrite Eq.~\ref{exactdiffeq} as an integral equation,
\begin{align}
\Psi(\mathbf{r}) & =\int \mathrm{d}  \mathbf{r}' \,G_0(\mathbf{r}-\mathbf{r}') \nonumber \\
& \quad \times  \left[ \sum_{i=1}^N \int \mathrm{d} \Omega_i \,\sigma_i(\mathbf{r}_i) \, \delta(\mathbf{r}'-\mathbf{R}_i) +s(\mathbf{r}') \right] , \label{regularpropsolution}
\end{align}
where we integrate over all positions $\mathbf{r}'$   ($\mathrm{d} \mathbf{r} \equiv \mathrm{d}^3 r$). In principle, we could choose the functions $\sigma_i(\mathbf{r}_i)$ to enforce the radiation boundary condition at each cell surface (Eq.~\ref{radiationBC2}), and then average over the cell positions  to obtain $\Sigma(\mathbf{r})$ in Eq.~\ref{linresponse}.   However, as discussed in Section~\ref{CellClusterUptake}, we do not have access to the full probability distribution $P(\{ \mathbf{R}_i^0 \}_{i=1}^N)$  necessary to do the averaging.  An alternative approach exploits an approximate solution which accounts for each cell independently to first order, the effects of pairs of cells at second order, etc.  This expansion requires knowledge of the probability  distributions of a single cell position, a pair of cells, a triplet, etc. Such scattering expansions are plagued with divergences due to the long-range nature of the diffusive interaction between the cells and must be treated via careful resummations  \cite{muthukumar2, kapral, cukier2}.  Particular care is required, because the Green's function, or propagator, in this expansion is $\hat{G}_0(\mathbf{q}) =  (D_{0 }q^2)^{-1}$ in Fourier space, which is singular as $q \rightarrow 0$.

To bypass such complications, we imagine cells imbedded in an ``effective medium'' with effective transport properties, to be determined self-consistently.  These transport properties  include screening  which cuts off the long range behavior of $G_0(r)$, and renders $\hat{G}(\mathbf{q})$ finite as $\mathbf{q} \rightarrow 0$. In particular, we assume a modified  ``effective medium'' propagator given by
\begin{align}
\hat{G}_{\Sigma}(\mathbf{q}) \equiv \frac{1}{D_0 q^2+\Sigma(\mathbf{q})}, \label{effectivemediumprop}
\end{align}
which describes the transport properties of the nutrient as it wanders through the homogeneous attenuating medium shown in Fig.~\ref{feffectivemedium}$(b)$.  A self-consistency condition on $\Sigma(\mathbf{q})$ leads to a renormalized (and better behaved) scattering expansion.

In Appendix~\ref{AEM}, we follow Cukier and Freed's effective medium calculation of  $\Sigma(\mathbf{q})$   \cite{cukier3}.   However, our cells will be partially absorbing, which generalizes the perfectly absorbing case Cukier and Freed  considered. The calculation assumes that the cells have uniformly distributed centers in the cluster.   Excluded volume interactions are included in an approximate way, by assuming that the  centers $\mathbf{R}_{i,j}^0$ of any pair of cells $i$ and $j$  are distributed according to the low density hard sphere pair distribution function
  \begin{align}
P(\mathbf{R}_i^0,\mathbf{R}_j^0) = \frac{1}{V^2} \, \theta( \left|\mathbf{R}_i^0-\mathbf{R}_j^0 \right|- 2a), \label{pairdistribution}
\end{align}
where $\theta(x)$ is the  step function and $V$ is the cluster volume.
  This generalization was also considered by Cukier \cite{cukier1}, but we find an important second order correction to his results from the pair distribution function Eq.~\ref{pairdistribution}.  The details are contained in  Appendix~\ref{AEM}.

 We find that the nutrient screening length in units of the cell radius $\alpha \equiv \xi/a$ satisfies
\begin{align}
\alpha^2 & =  \frac{3 \phi\alpha  \nu(1+\coth \alpha)   }{1+ \alpha +\nu}+36\phi^2 \left[\frac{  \alpha\nu(1+\coth \alpha)   }{1+ \alpha +\nu} \right]^2\nonumber  \\
& \quad   \times   \left[  \frac{1}{4 \alpha^2}+k_1(2\alpha)i_0(\alpha) \left[  \frac{ \alpha i_1(\alpha) }{\nu}-i_0(\alpha) \right] \right] \label{finalxisolution},
\end{align}
where $i_{\ell}(x)$ and $k_{\ell}(x)$ are the modified spherical Bessel functions of the first and second kind, respectively.  For a specific $\nu = \kappa a$, Eq.~\ref{finalxisolution}  is a self-consistent equation for $\alpha=\xi/a$. It is soluble numerically (we used Newton's method of successive approximation) for $\xi$ as a function of $\nu$ and $\phi$.

\begin{figure}
\includegraphics[height=2in]{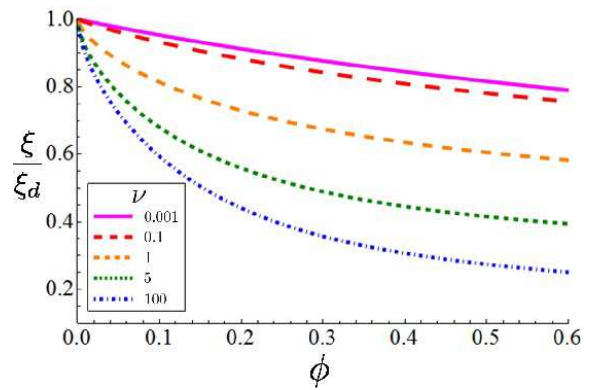}
\caption{\label{xiEMfig} (Color online) Plot of the  screening length $\xi$ calculated with the effective medium theory divided by the dilute limit result $\xi_d$ as a function of the volume fraction $\phi$ for various values of $\nu$. The low density result becomes increasingly inaccurate at larger packing fractions.}
\end{figure}

The ratio $\xi/\xi_d$, where $\xi_d$ is the dilute limit formula Eq.~\ref{dilutexi}, appears as a function of $\phi$ for various  $\nu$ in Fig.~\ref{xiEMfig}.  Note that the effective medium theory always predicts  $\xi < \xi_d$, a plausible result since this more sophisticated theory allows for  nutrient absorption mediated by repeated interactions with individual cells.
For the values of $\nu$ typical of yeast clusters and gram negative bacteria ($\nu \sim 0.001-0.1$), $\xi$ is no more than 20\% different from $\xi_d$ even for large $\phi \sim 0.5$.  These results suggest that effective medium theory is a reasonable approximation for dense clusters of cells with $\nu = \kappa a \ll 1$, in contrast to the more problematic perfectly absorbing case $\nu \rightarrow \infty$    \footnote{The effective medium calculation in the perfectly absorbing $\nu \rightarrow \infty$ limit has serious deficiencies: even the second order correction calculated in Appendix~\ref{AEM} yields unphysical results for large values of $\phi$ \cite{cukier3}. }. 

\begin{figure}
\includegraphics[height=2in]{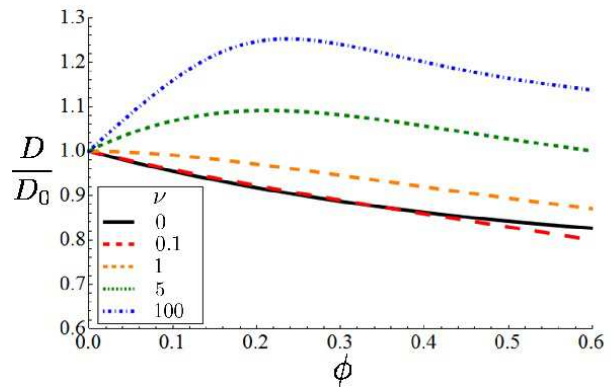}
\caption{\label{DEMfig} (Color online) Plot of the  diffusion coefficient $D$ in the cell cluster calculated with the effective medium theory divided by the bare diffusion coefficient $D_0$ in the absence of the cluster as a function of the volume fraction $\phi$ for various values of $\nu$. }
\end{figure}

The second order term in the Taylor series expansion of $\Sigma(\mathbf{q})$ around $\mathbf{q}=0$ yields an implicit equation for the effective diffusion coefficient $D$ of the nutrient inside the cell cluster.  The somewhat cumbersome equation is Eq.~\ref{solutionfordiff} in Appendix~\ref{AEM}.  The solutions for various $\nu$ are plotted in Fig.~\ref{DEMfig} as a function of $\phi$.  In all cases of biological interest, the corrections to the bare diffusion coefficient $D_0$  in the absence of the cluster are relatively small, i.e., approximately 10-20\% and never exceeding 30\%.

 In the biologically relevant regime $\nu \lesssim 0.1$, the correction to the diffusion coefficient is close to the $\nu=0$ (perfectly reflecting) limiting case given by 
\begin{equation}
\frac{D(\nu=0)}{D_0} = \frac{1}{1+\phi/2-\phi^2/4}, \label{nu0D}
\end{equation}
which is derived from  Eq.~\ref{solutionfordiff}.  As expected, the diffusion coefficient $D$ is smaller than $D_0$  when the cells are perfectly reflecting.  Also, the structure of  Eq.~\ref{nu0D} is consistent with the modified Maxwell-Garnett theory for diffusion of particles in a medium with perfectly reflecting inclusions derived in Ref.~\cite{maier2002}.
  Eq.~\ref{nu0D} is also consistent with experimental results for the diffusion coefficient of water in a suspension of spherical colloidal particles \cite{linse1986}.  At high volume fractions, we expect the effective medium theory to start to break down and higher order terms in $\Sigma(\mathbf{q})$ to contribute to transport within the colony.  Thus, the curves in Fig.~\ref{DEMfig} should be treated as approximations and specific features, such as the crossover between the $\nu=0$ and $\nu=0.1$ lines, might not be physical.

   We now check the effective medium result against simulations and experiments.

\section{\label{SSimulations}Numerical Simulations}

To obtain further insight into nutrient absorption, we now go beyond effective medium theory and solve the diffusion equation numerically in a densely randomly packed cluster of spherical cells.  We employ the finite element solver discussed below to numerically solve the steady-state diffusion equation with the appropriate boundary conditions.  Given the complicated nature of the problem, we can only test effective medium theory for clusters of up to 400 cells in this way.  However, since the effective medium approximation is derived in the limit of an infinite number of cells, if it works for simulations with 100-400 cells, it should be even more reliable when we have even more cells, as in a typical growing yeast colony.

\subsection{\label{SSimulIntro}Cell Cluster Simulation}

To check the analytic results, we solved the steady state diffusion equation exactly for  cell clusters with hundreds of cells.  The numerical solution was found with the COMSOL 3.5a finite element solver \cite{COMSOL}.  A MATLAB program was written to input in the locations and radii of all the cells in the cluster.  The COMSOL program included a computer-assisted design (CAD) feature that was then able to parse the MATLAB output and create a particular arrangement of spherical cells that defined our domain of interest.  The coordinate list for the sphere cluster was created via a Bennett model, originally designed to quickly generate amorphous, dense random packings of identical spheres \cite{bennett}.  These arrangements approximate the disordered packing of cells observed in yeast colonies in the experiments.

\begin{figure}
\includegraphics[height=1.7in]{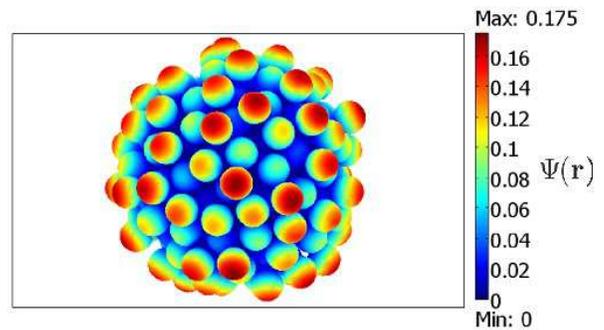}
\caption{\label{fcluster}(Color online) A simulated cluster of cells with $\nu = 1$ and $\psi_{\infty} = 1$.  The color (shading) indicates the local nutrient concentration $\Psi(\mathbf{r})$ near the cell surfaces. In this case, the 200 cells with radius $a$ were confined to a cluster of radius $b \approx 7.45 a$, yielding a volume fraction of $\phi \approx 0.48$.  The radius of the large bounding sphere on which the nutrient concentration was fixed at $\psi_{\infty}$ was about $56a \approx7.5b $.  To better simulate the fixed concentration $\psi_{\infty}$ infinitely far from the cluster,  we used COMSOL's ``infinite element'' option with spherical symmetry between radii $28a$ and $56a$ (see Sec.~\ref{SSimulIntro}).}
\end{figure}

To prevent the cells from touching and creating singularities in the finite element mesh, identical spherical cells with radius $a$ are placed  at the sphere centers of a cluster generated by the Bennett model using spheres with larger radius  $\tilde{a}>a$.  This guarantees a gap of at least $2(\tilde{a}-a)$ between adjacent spheres.  A high volume fraction $\phi \approx 0.63$ (corresponding to random close packing) is generated with  $\tilde{a}=1$ and $a = 0.999$.   Smaller values of $\phi$ are generated by decreasing $a$  for a fixed $\tilde{a}=1$.  For example, $a=0.9$ results in a volume fraction $\phi \approx 0.48$ (e.g., the cluster in Fig.~\ref{fcluster}).    

A large bounding sphere concentric with the center of mass of the cell cluster allowed us to impose a constant nutrient concentration ``at infinity'':  $\Psi(|\mathbf{r}| \rightarrow \infty) = \psi_{\infty}$, to approximate a suspended cell cluster in an infinite nutrient bath buffered at concentration $\psi_{\infty}$.
We used COMSOL's  ``infinite element'' option to efficiently simulate by placing extremely large finite elements between the bounding sphere and the cell cluster.   With bounding spheres of radii typically  five times the  cell cluster radius, this method yielded good results: Our calculated solution was unchanged when varying the bounding sphere radius from 5 to 10 times the cluster radius.  

 Finally, we specified the dimensionless nutrient uptake parameter  $\nu=\kappa a$ for cells with identical radius $a$ and applied the boundary conditions given by Eq.~\ref{radiationBC2} on each cell surface.  A particular cell cluster and the corresponding steady state nutrient concentration $\Psi(\mathbf{r})$ near each cell surface is shown in Fig.~\ref{fcluster}.  In this case,  $\nu = 1$, the packing fraction $\phi \approx 0.48$, and the cluster radius $b \approx 7.45a$, with $\psi_{\infty} = 1$.  The bounding sphere had a radius of $56a$ and we inserted infinite elements at distances between $28a$ and $56 a$.   With these parameters, the effective medium theory predicts a screening length $\xi \approx 0.72 a$.  This is consistent with what we observe in Fig.~\ref{fcluster}, as the concentration decays by a factor of $e$ over a distance comparable to the cell radius $a$.

\subsection{Comparison with Theoretical Results}

\begin{figure}
\includegraphics[height=2in]{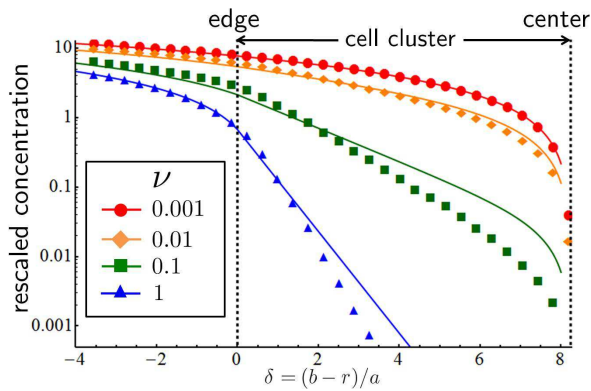}
\caption{\label{clusterdecayfig}    (Color online) The  radially averaged  concentration $r\overline{\Psi}(r)/a \psi_{\infty}$ (symbols) and the  effective medium result  $r \psi( r)/a \psi_{\infty}$ (lines)  for a   cluster of  353 cells of radius $a$ (with cluster radius $b \approx 8.15a$) as a function of $\delta =(b-r)/a$ on a log-linear plot. The concentrations are rescaled to highlight the exponential decay of the concentration into the bulk of the cluster. Cells occupy the region $0 < \delta  < 8.15$ with the  center at $\delta  \approx 8.15$,  as indicated by the dashed lines.  The cell packing fraction is $\phi \approx 0.63$.  The bounding sphere has a radius of $50a$, with ``infinite elements''  inserted at distances between $25a$ and $50a$ (see  Sec.~\ref{SSimulIntro}).   }
\end{figure}

To compare theoretical and simulation results, we tracked the decay of the radially averaged nutrient concentration $\overline{\Psi} (r) \equiv  (4 \pi)^{-1} \int \Psi(\mathbf{r})    \, \mathrm{d} \Omega$ into the center of the colony,  and compared with the effective medium prediction $\psi(r)$ of Eq.~\ref{psisol}.  Fig.~\ref{clusterdecayfig} shows a semilogarithmic plot of $r \psi(r)/a\psi_{\infty}$ versus $\delta = (b-r)/a$, varying $\nu = \kappa a$ over 3 orders of magnitude.  Eq.~\ref{psisol} predicts that this quantity  decreases exponentially near the cell surface when $1 \ll b/a$.  Fig.~\ref{clusterdecayfig} shows how the nutrient concentration decreases as we move into the cluster.

\begin{figure}
\includegraphics[height=2in]{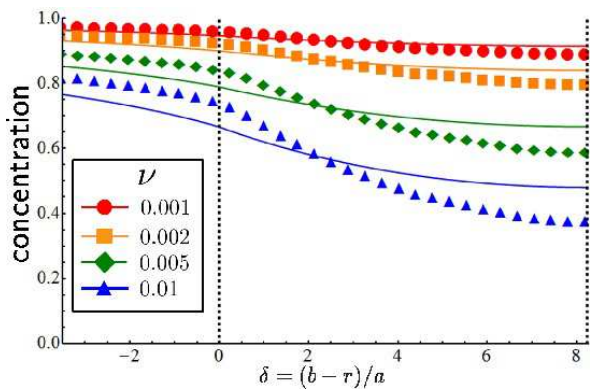}
\caption{\label{lownufig} (Color online) The radially averaged  concentration $\overline{\Psi}(r)/\psi_{\infty}$ (symbols) and the effective medium solution  $\psi(r)/\psi_{\infty}$ (lines) as a function of $\delta=(b-r)/a$ for the same 353 cell cluster described in Fig.~\ref{clusterdecayfig}.  The dotted lines indicate the cluster edge ($\delta = 0$) and center ($\delta  \approx 8.15$). }
\end{figure}

Fig.~\ref{clusterdecayfig} shows that, even for the large volume fraction $\phi \approx 0.63$, effective medium theory provides an excellent description, especially for small values of $\nu$.  Thus, this theory is appropriate for modelling yeast colonies, which have a very small $\nu \sim 0.001$ value.       The effective medium agreement should improve if computer resources allow more cells in the simulation, since it was designed to handle the limit where the cell number $N \rightarrow \infty$ and cluster volume $V \rightarrow \infty$, with $N/V$ fixed.
    Fig.~\ref{lownufig} shows the low $\nu$ regime (poor nutrient absorbtion) in more detail. Although the differences between the simulation and effective medium theory are now more evident, the absolute difference between the simulation and the theory for the concentration remains small.

\begin{figure}
\includegraphics[height=2.6in]{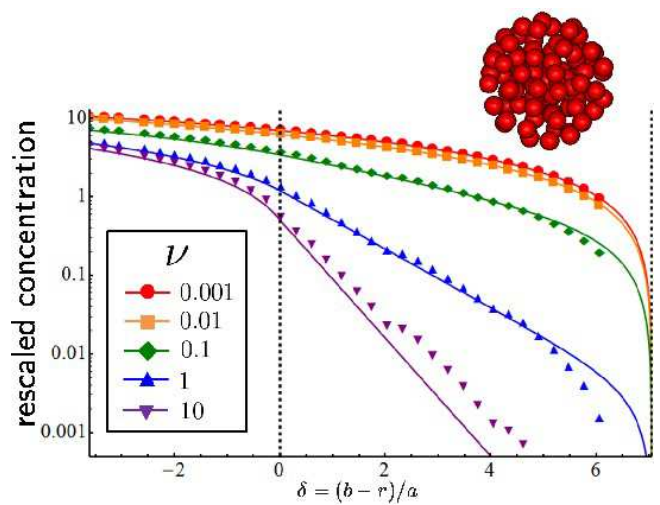}
\caption{\label{diluteclusterfig} (Color online) Exponential decay of the rescaled nutrient concentration in the effective medium theory  (lines) and  in a simulation (symbols) for a more dilute 85 cell cluster with $b = 7a$ and various $\nu$. The  cell packing fraction is $\phi \approx 0.24$.  The cluster, shown above the graph, now occupies the region between the dotted lines, $0 < \delta < 7$.  }
\end{figure}
We also studied a more dilute 85 cell cluster, to check that effective medium theory is indeed accurate for lower volume fractions and an alternative cell configuration.  Instead of using the Bennett model, we  placed cells with centers in three concentric shells with radii   $2a$, $4a$ and $6a$.  The cells within each shell were placed randomly, but their positions were adjusted to prevent the cells from touching and disrupting the finite element mesh. Fig.~\ref{diluteclusterfig} reveals even better agreement between the simulation and effective medium theory for the lower volume fraction of $\phi \approx 0.24$.  Note that good agreement in this case is obtained for the highly absorbing limit $\nu \geq 1$, as well.  We conclude that effective medium theory provides a good description of spherically averaged nutrient uptake, for both weakly and highly absorbing cells,  for volume fractions $\phi \lesssim 0.6$.

\section{\label{SExperiments}Experiments}

 We now compare the effective medium theory presented in Sec.~\ref{STheory} to experimental results for glucose uptake in yeast cell colonies.   Yeast cell aggregates are a particularly interesting biological application because they form naturally in the wild and are a possible model for the initial emergence of multicellularity \cite{Koschwanez2011}.  Nutrient uptake in single versus clumped cells is a crucial factor in this model system.  Thus, accurately modelling  nutrient transport properties in cell clusters is key to understanding their biological function and role in evolution.

Since yeast cells are poor glucose absorbers, the characteristic nutrient penetration depth in a yeast colony is much larger than a cell radius and is easily visible in colony cross sections, as shown in Fig.~\ref{fconfocal}. Their poor absorption properties are also well-suited for effective medium modelling, since we expect that nutrient transport for small absorption parameter $\nu \ll 1$ is dominated by much longer length scales than the local colony geometry at the single cell scale.

\subsection{\label{SExpSetup}Experimental Setup}

\begin{figure}
\includegraphics[width=3in]{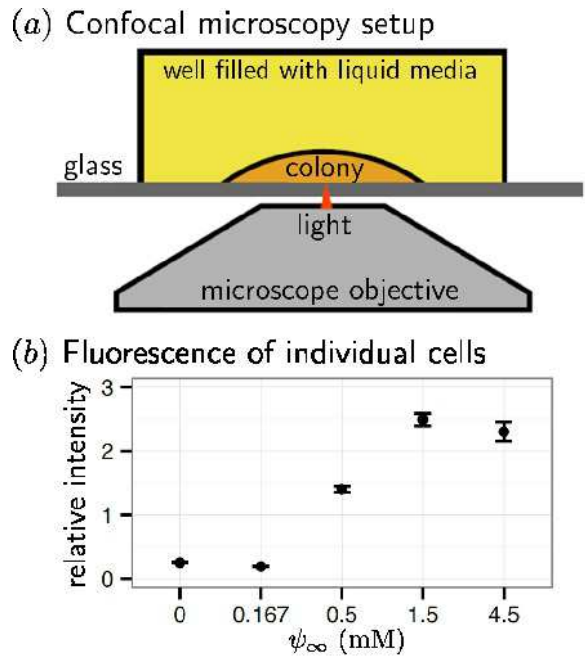}
\caption{\label{expsetupfig} (Color online) $(a)$ Confocal microscopy setup. A single yeast cell was inoculated into a well of a glass-bottomed 96-well plate containing 200 $\mu$L of yeast media and allowed to grow into a colony. The bottom of the colony was imaged using a spinning-disc confocal inverted microscope. $(b)$ Cells were grown in shaken media, pipetted into an empty well, and imaged immediately (before they could divide and form colonies) with the setup shown in $(a)$. At least 250 cells at each concentration were imaged, and the error bars are the 95\% confidence interval using the Student's $t$-test. Relative fluorescence intensity  in $(b)$ is defined as ribosomal protein expression divided by constitutive expression.  Cells in 0.167 mM glucose continue to divide, but at less than half
the growth rate of cells in 0.5 mM glucose. }
\end{figure}

We used confocal microscopy to measure the distribution of growth in a yeast colony. To start the experiment, we filled each well of a 96-well, glass-bottomed microtiter dish (Greiner Bio-One 655892) with 200 $\mu$L of minimal, synthetic yeast media (recipe in Ref.~\cite{Koschwanez2011}). Figure~\ref{expsetupfig}$(a)$ shows the confocal setup and Fig.~\ref{fconfocal} shows representative images.  We engineered a yeast strain (yJHK192) with two fluorescence protein (FP) reporters: (1) Cerulean, a stable  CFP (cyan) variant, was constitutively expressed from the \textit{ACT1} promoter; and (2) mCitrine, a YFP (yellow) variant, was expressed from the ribosomal protein  \textit{RPL7A} promoter. The expression of  \textit{RPL7A} has been shown to correlate with growth rate \cite{Brauer}; the correlation between growth rate and glucose concentration has been well studied in yeast \cite{Barnett}, and we verified that the YFP fluorescence correlated with glucose concentration in cells grown in shaken media (Figure~\ref{expsetupfig}$(b)$). mCitrine was destabilized by adding a \textit{CLN2} PEST-sequence to the C terminus which is expected to yield a protein half-life of about 30 minutes \cite{Mateus}; a destabilized fluorescent protein was essential to see decreased expression of \textit{RPL7A} in cells that had stopped growing.   Note that the 30 minute half-life is much shorter than the yeast doubling time of 2.5 to 3.5 hours in low glucose.

When grown on agar, small yeast cell colonies  (less than 1 mm across) form a spherical cap with a contact angle $\theta$ between the edge and the agar surface.   The contact angle $\theta$ increases with agar concentration and $\theta \approx 40^{\circ}$ for 2\% agar \cite{colonyshape}.  Unfortunately, in our experiment, the side view of the colonies was optically inaccessible  and a precise characterization of the colony morphology was not possible.  However, since our colonies are very small, we expect that surface energy minimization is most important in determining the shape so that the colonies are approximately spherical caps. Also, we know the edges of the colonies were at
least three cells thick (the maximum depth we were able to see by
microscope).

We grew colonies in three glucose concentrations: 0.5 mM, 1.5 mM, and 4.5 mM. We inoculated one cell per well using a fluorescent activated cell sorter (MoFlo FACS, Beckman Coulter, Inc.); the cells had been grown to saturation in 1 mM glucose synthetic media. Inoculation was verified under a microscope, and those wells with a cell closest to the middle of the well (at least 3 wells per glucose concentration) were selected for colony tracking. The plate was incubated without shaking at 25 ${}^{\circ}$C for 3 days. Each colony was imaged a day after inoculation and an additional 2-3 times over the next 2 days.  The radius of the largest imaged colony in each glucose concentration
was less than half of the average colony size after two weeks of
growth (610 $\mu$m in 0.5 mM glucose, 920 $\mu$m in 1.5 mM glucose, and 1170
 $\mu$m in 4.5 mM glucose), indicating that the carbon source was not yet
depleted. Images were taken with a 20X objective on a Nikon inverted Ti microscope  with a Yokagawa spinning disc unit  and an EM-CCD camera (Hamamatsu ImagEM); CFP was excited with a 447 nm laser and YFP was excited with a 515 nm laser; exposure times for all images (including single-cell images) were 200 ms (CFP) and 1000 ms (YFP). All images were focused on the bottom layer of cells in the colony, and multiple, overlapping images were taken of colonies that exceeded the field of view.  Three independent experiments were performed. Figures \ref{expconcdecayfig} and \ref{pendepthfig} show the combined results of all experiments.

\subsection{\label{Yeast} Data Analysis and Results}

\begin{figure}
\includegraphics[width=3.4in]{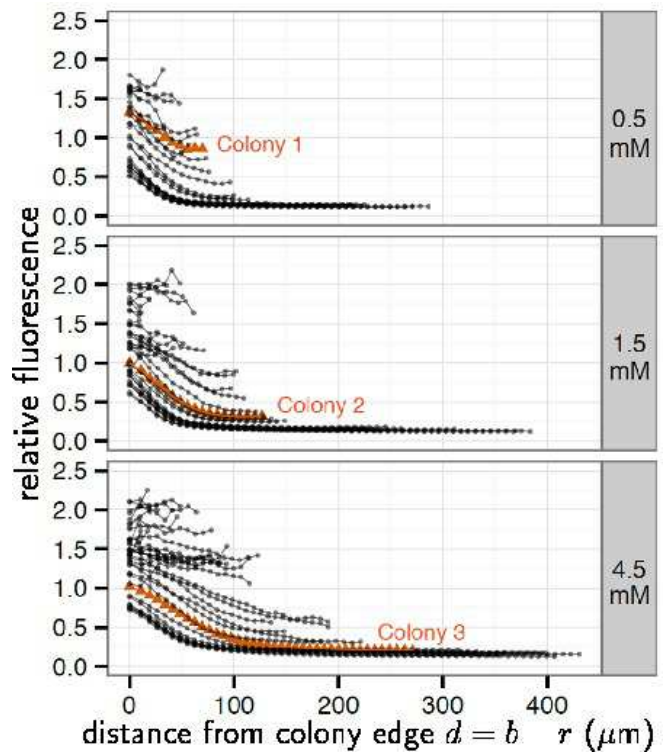}
\caption{\label{expconcdecayfig}  (Color online) Relative fluorescence intensity (ribosomal protein expression divided by constitutive expression) as a function of distance $d=b-r$  from the edge of the colony.  Here $b$ is the colony radius and $r$ is the observation position. Each point is the average fluorescence of an 8 $\mu$m thick ring whose edge is located an equal distance from the edge of the colony. The last point on each connected set of lines represents the radius of the colony and larger radii reflect longer periods of incubation. Each line is one colony. Colonies were grown in 0.5 mM glucose (top), 1.5 mM glucose (middle), and 4.5 mM glucose (bottom). Profiles from the three colonies shown in Fig.~\ref{fconfocal} are shown as orange (gray) triangles. }
\end{figure}

Images were processed using the Fiji distribution of ImageJ \cite{Schindelin}. Images were converted to 8-bit, stitched together \cite{Preibisch}, and merged into a single RGB image. Contrast was not adjusted during processing. Fluorescence intensity as a function of colony radius was measured using a custom script written in Python. The basic algorithm is as follows: (1) The constitutive (CFP) image was thresholded using Li's Minimum Cross Entropy thresholding method \cite{LiTam}. (2) Noise and cells not attached to the colony were removed by eroding and then dilating the binary image a total of three times with a four pixel diameter circular structuring element. (3) A series of mask "rings" was made by a series of morphological erosions and subtractions from the original image. The distance between rings was approximately 8 $\mu$m, or 2 cell diameters. (4) The masks were used to generate images of concentric rings using both the constitutive (CFP) and the growth-dependent expression (YFP) image. (5) The average fluorescence of each ring was measured; the reported, relative fluorescence is the average expression (YFP) fluorescence divided by the average constitutive (CFP) fluorescence. Figure~\ref{expconcdecayfig} shows the intensity as a function of distance from the edge of the colony; each point in the figure is a measurement of fluorescence at one of the concentric rings, and a line joins the values from a single colony.

In order to compare the experimental results to the effective medium theory, we assume that each colony is a hemisphere with a radius $b$ equal to the radius of the bottom cross-section.   Since the colonies are  small compared to the glass well dimensions and since the glucose molecules do not stick to the glass, we also assume that the hemispheres sit on an infinite, reflecting surface.    At the infinite surface $I$, the perfectly reflecting boundary condition holds: $ \left. \hat{\mathbf{n}} \cdot \nabla \psi (\mathbf{r})\right|_{\mathbf{r} \in I} = 0$, where $\hat{\mathbf{n}}$ is the surface normal.  Also, note that we can bisect a spherical colony in an infinite medium into two hemispheres with an imaginary plane.  The  symmetry of the spherical colony across the plane guarantees that the same boundary condition $ \left. \hat{\mathbf{n}} \cdot \nabla \psi (\mathbf{r})\right|_{\mathbf{r} \in I} = 0$ holds on this imaginary surface, as well.  Therefore, the region on one side of the imaginary plane has the same boundary conditions as a hemispherical colony sitting on an infinite reflecting plane.  Hence, the concentration profile $\psi(r)$ for a hemispherical colony on a reflecting surface is given by the solution for spherical colonies (Eq.~\ref{psisol}), where $r$ is the distance from the colony center (measured above the infinite plane).

As discussed in the previous section, the colonies in the experiment could be spherical caps with a smaller contact angle $\theta < 90^{\circ}$ at the glass well bottom.   A smaller contact angle corresponds to a more shallow colony with an increased nutrient penetration depth, due to nutrient diffusion from the top of the colony.   However, we expect that the corrections due to smaller $\theta$ to be small when the penetration depth $\ell$ is small compared to the colony height.

Some predictions of the effective medium theory are qualitatively confirmed by the experimental data. The effective medium theory described in Sec.~\ref{STheory} predicts that for these dense cell colonies, the characteristic shielding length is 50 $\mu$m, or about 12 cell diameters.  This is consistent with the fluorescence curves in Fig.~\ref{expconcdecayfig}.  The theory also predicts that the glucose level at the outer edge of the colony decreases like $1/b$, where $b$ is the colony radius (see Eq.~\ref{psisol}). We see this effect in Fig.~\ref{expconcdecayfig}, where shorter curves (corresponding to smaller colonies) have a higher fluorescence level at $d= 0$. However, the fluorescence level has a complicated, non-linear relationship to the glucose concentration in the bulk medium (see Fig.~\ref{expsetupfig}$(b)$) and the specific $1/b$ scaling cannot be tested.

\begin{figure}
\includegraphics[height=1.7in]{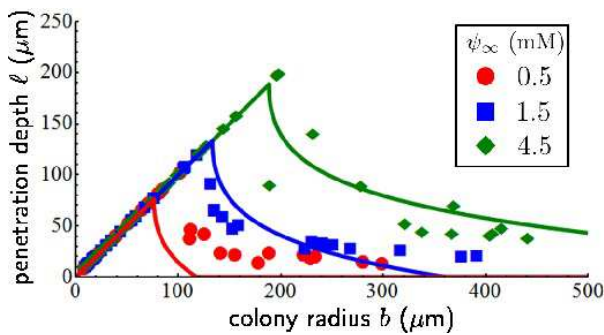}
\caption{\label{pendepthfig}    (Color online) Penetration depth $\ell$ as a function of colony radius $b$. Penetration depth in the experiment (symbols) is the distance from the edge of the colony at which relative fluorescence drops below 0.4.  The theoretical results (solid lines) are calculated by numerically solving $\psi(r=b-\ell)=\psi_{\mathrm{min}}$ (Eq.~\ref{psisol}) for   $\ell$, where we assume that $\psi_{\mathrm{min}} \approx 0.25$~mM local glucose concentration corresponds to a 0.4 relative fluorescence level.  As discussed in the text, this level is slightly higher than the level at which the cells stop growing and expressing the ribosomal protein. We also use a cell radius of $a = 2$~$\mu$m, a cell packing fraction $\phi = 0.56$, and  $\nu = 6 \times 10^{-4}$.     }
\end{figure}

Figure~\ref{pendepthfig} shows the penetration distance $\ell$ as a function of colony size; penetration distance is defined here as the distance from the edge of the colony at which the relative fluorescence intensity falls to less than 0.4.  This intensity is chosen to be well above the background fluorescence level.  For the smaller colonies, the fluorescence is higher than 0.4 throughout the colony. For these colonies, we say that the penetration depth is equal to the colony radius: $\ell = b$.  Eq.~\ref{psisol} is used to find a theoretical result for the penetration depth $\ell$ inside the colony.  We assume that the relative fluorescence intensity of 0.4 corresponds to a local glucose concentration of $\psi_{\mathrm{min}} \approx 0.25$~mM.  Again, note that this level is slightly higher than the minimum glucose level required for growth.  We estimate the packing fraction of cells inside a colony, $\phi \approx 0.56$, by looking at cell colonies growing on agar (data not shown).  Then, assuming a cell radius $a = 2$~$\mu$m  and absorption parameter $\nu = 6 \times 10^{-4}$, we find reasonable agreement between the theoretical solution for $\ell$ and the experimental data in Fig.~\ref{pendepthfig}.

 Assuming a hemispherical colony, it is also possible to approximate the total number  $N$  of growing cells  inside the yeast colony.  If $\ell$ is the penetration depth, then $N \approx 2 \pi \phi (b^3-(b-\ell)^3)/3$.   In the experiment, a direct measurement of $N$ is not possible because we only have access to a single slice through the colony. However,  from the penetration depth $\ell$ results in Fig.~\ref{pendepthfig}, we see that for small colonies, $\ell = b$ and all of the cells grow so that $N \propto b^3$.  As the colony gets larger, the center stops growing and the penetration depth  $\ell$ starts to decrease.    An analysis of Eq.~\ref{ellatlargeb}  predicts that $\ell$ is constant, but with negative contributions that grow logarithmically with $b$.       Hence, $N$ should grow like $b^2$  with logarithmic corrections.   The experimental results in Fig.~\ref{pendepthfig} have the predicted decrease in $\ell$ at large radii $b$. However, a test of the specific scaling arguments for $N$  would require testing larger colonies and measuring the three dimensional colony shape.

We do see some disagreements between the experimental results in Fig.~\ref{pendepthfig}, especially at the 0.5 mM glucose concentration.   In general, we expect to underestimate the penetration depth due to our approximation of colonies as hemispheres. The deviations may also indicate the presence of biological mechanisms that allow the cell clusters to adapt to particular nutrient environments.  For example, yeast cells can modify their nutrient uptake parameter $\nu = \kappa a$  by expressing different nutrient transporters on their surface \cite{difftrans}.  Our theory assumes that each cell in a colony has the same absorption strength $\nu$, an  approximation suitable provided all cells employ the same nutrient transporter system, and the ambient nutrient concentration is small enough so that we are in a linear regime of nutrient absorption in a Michaelis-Menten-type curve (see Fig.~\ref{MMfig}).  So, deviations from our theory may indicate deviations from the linear regime or the presence of regulatory mechanisms in cells that modify nutrient uptake kinetics during colony growth.

\section{\label{SConclusions}Conclusion}

We have developed and tested an effective medium theory of nutrient transport in clusters of cells.   Simulations and experiments support the theoretical results.  The key parameters in our model are the single cell nutrient absorption parameter $\nu$ and the packing fraction $\phi$ of cells in the cluster.  The parameter $\nu$ can vary widely depending on many factors such as the nutrient transporter expression, transporter kinetics, and cell wall thickness.  We expect $\nu$ to decrease with increasing cell wall thickness as discussed in Sec.~\ref{SingleCell}.   

The wide variation in  $\nu$ for different cell types such as yeast $(\nu \sim 0.001)$, gram-positive $(\nu \sim 0.01$), and gram-negative $(\nu \sim 0.1$) bacteria
has important  implications for  nutrient absorption.  Effective medium theory predicts very different nutrient shielding properties of cell clusters as $\nu$ varies (see Figs.~\ref{clusterdecayfig} and  \ref{diluteclusterfig}).  Thus, the fraction of actively growing cells at a cluster surface will  vary significantly with cell types.  For instance, we predict that the screening length $\ell$ for an \textit{S. cerevisiae} colony is about 10 times longer than the length in a gram negative  \textit{E. coli} bacteria colony and 3 times longer than in a gram positive  \textit{L. mesenteroides} bacteria colony, measured in units of the respective cell diameters.   More specifically,  the glucose concentration in a tightly packed yeast colony ($\phi\approx 0.56$) will fall off exponentially into the colony with a characteristic ($e$-folding) length of 50 $\mu$m.  A similarly packed \textit{E. coli} colony will have a much more rapid falloff with a 1 $\mu$m characteristic length.    

Our theory also predicts that nutrient shielding is  more sensitive to the volume fraction $\phi$ when each cell is a good absorber (i.e., for $\nu \gtrsim 0.1$). As $\phi$ increases, correlations between the cells become more important as nutrient collisions with multiple cells create a stronger shielding effect.    Thus, a colony of good absorbers, such as  gram-negative bacteria, should be able to change its nutrient shielding properties by tuning the separation between each cell.  This hypothesis could be tested by placing  bacterial cells in a gelatinous matrix at various cell densities and observing their growth. As discussed in Sec.~\ref{SSDiluteCells}, these artificial colonies have many industrial and environmental applications. 

 It would be interesting to extend our theory to include nonspherical cell shapes and cell radius polydispersivity.  Polydispersivity should increase the screening length, but a detailed understanding of its effect on cell correlations is lacking.  One could also account for spatial variability in $\phi$.  Cell clusters should ``thin out'' near the cluster surface, where the cells have not had time to grow into a densely packed structure. If this density variation occurs on scales large compared to the cell radii, it should be sufficient to replace $\phi$ by a spatially dependent $\phi(\mathbf{r})$ in our effective medium calculations.   

 To extend the experimental results and test the effective medium theory more precisely, one could control for the variability of the $\nu$ parameter in the colony.  One possibility is to use engineered yeast strains with fixed nutrient uptake kinetics, such as the mutants constructed by Reifenberger et al. \cite{transyeast}.  These cells  express a single type of glucose transporter in a  medium with a low glucose concentration.  In addition, it would be interesting to more precisely characterize the local glucose concentration in the colony  by using either a more direct reporter or more precisely characterising the relationship between glucose level and ribosomal protein expression.

\begin{acknowledgments}

We thank Andrew W. Murray for helpful discussions, generous advice, and comments on the manuscript. MOL acknowledges the support of the National Science Foundation Graduate Research Fellowship.  Experimental work by JHK and MOL in the Andrew Murray Lab was supported by National Institute of General Medical Sciences Center of Excellence grant P50 GM 068763 of the National Centers for Systems Biology.   Theoretical work by DRN and MOL was supported by this grant also, and as well as by the National Science Foundation through grant DMR-0654191 and the Harvard Materials Research Science and Engineering Center through grant DMR-0820484. 
  Portions of this research were done at the Kavli Institute for Theoretical Physics at Santa Barbara,  supported in part by the National Science Foundation under Grant No. PHY11-25915.
\end{acknowledgments}

\appendix

\section{\label{AEM}Detailed Effective Medium Calculation}

Our effective medium calculation of the response function $\Sigma$ will closely follow Cukier and Freed's analysis of the perfectly absorbing cell case \cite{cukier3}.  We generalize their work by allowing for partially absorbing cells via the radiation boundary condition (Eq.~\ref{radiationBC2}).  Cukier also considered this case \cite{cukier1}, but only to first order in scattering in the effective medium.  We  extend his argument to include the pair distribution function of the cells, which leads to an important correction term.

We begin by solving for the Lagrange multiplier functions $\sigma_{\alpha}(\mathbf{r}_i)$ in Eq.~\ref{exactdiffeq} in terms of  an ``effective medium propagator'' $G_W(\mathbf{r},\mathbf{r}')$. Upon assuming the effective nutrient transport properties are homogeneous over the cell cluster, the propagator is translationally invariant ($G_W(\mathbf{r},\mathbf{r}') \equiv G_W(\mathbf{r}-\mathbf{r}')$) and, in Fourier space,
\begin{equation}
\hat{G}_{W}(\mathbf{q}) \equiv \left[  D_0q^2 + W(q) \right]^{-1}, \label{GWFT}
\end{equation}
 where $W$ is some suitable approximation to  the self-energy $\Sigma$ that tells us how nutrients diffuse through a medium consisting of the nutrient sinks and their interstitial space..  To simplify the analysis, we introduce an operator notation for all of our convolution integrals.  For example,  $[W \Psi]_{\mathbf{r}} \equiv \int W(\mathbf{r}-\mathbf{r}')\Psi(\mathbf{r}') \, \mathrm{d} \mathbf{r}'$.  Subtracting the convolution $W \Psi$ from both sides of the equation for $\Psi(\mathbf{r})$ (Eq.~\ref{exactdiffeq}) yields
\begin{align}
\Psi (\mathbf{r}) & = -\int \mathrm{d} \mathbf{r}'\,G_W(\mathbf{r}-\mathbf{r}') \nonumber \\
& \qquad \qquad \qquad \times   \left[ \sum_{i=1}^N  \tilde{\sigma}_i(\mathbf{r}')+s(\mathbf{r}')-[W \Psi]_{\mathbf{r}'} \right]\,  \nonumber \\
& \equiv -G_W\left[ \sum_{i=1}^N  \tilde{\sigma}_i+s-W \Psi \right] , \label{exactsolution}
\end{align}
where $ \tilde{\sigma}_i (\mathbf{r})= \int \mathrm{d} \Omega_i \, \delta(\mathbf{r} - \mathbf{R}_i) \sigma_i(\mathbf{r}_i)$, and $G_W \left[ \ldots \right] \equiv \int G_W(\mathbf{r}-\mathbf{r}') [ \ldots ]_{\mathbf{r}'} \, \mathrm{d} \mathbf{r}'$ is another example of the convolution described above.  

To solve for  $\sigma_i$ via Eq.~\ref{exactsolution},  denote the inverse of the $G_W$ operator on the $i^{\mathrm{th}}$ cell surface by $K_{i}^{-1}(\mathbf{r}_i,\mathbf{r}_i')$, where $\mathbf{r}_i$ and $\mathbf{r}'_i$ are two vectors from the cell origin to the cell surface (see Fig.~\ref{feffectivemedium}$(c)$).     Then,  since Eq.~\ref{exactsolution} is an equation for $\Psi(\mathbf{r})$ for all points $\mathbf{r}$ in the cluster,  $\Psi(\mathbf{r})$ in Eq.~\ref{exactsolution} is evaluated at an arbitrary point $\mathbf{r}=\mathbf{R}_i$ on the $i$-th cell surface.
 In the perfectly absorbing case Cukier and Freed \cite{cukier3} considered, the boundary condition $\Psi(\mathbf{R}_i)=0$ is then used to solve Eq.~\ref{exactsolution} for $\sigma_i(\mathbf{r}_i)$.   The analogous condition for  partially absorbing cells  is  Eq.~\ref{radiationBC2}. Using the partially absorbing boundary condition yields 
\begin{align}
 \sigma_i & (\mathbf{r}_i)=- \int  \mathrm{d} \Omega_i' \, g_i(\mathbf{r}_i,\mathbf{r}_i')  \nonumber \\
 & \qquad  \quad  \times  \left[  (1-\mathcal{Q}_i')G_W \left(s-W \Psi   + \sum_{j \neq i}  \tilde{ \sigma}_j\right) \right]_{\mathbf{r'}=\mathbf{R}_i'}. \label{sigmasol2}
\end{align}
where   $g_i \equiv \left[1- K_i^{-1}\mathcal{Q}_i G_W  \right]^{-1} K_i^{-1} $ is an operator defined on the surface of the $i^{\mathrm{th}}$ sphere and does not depend explicitly on the sphere center $\mathbf{R}_i^0$. $\mathcal{Q}_i'$ is the gradient operator with respect to the $\mathbf{r}'$ coordinate. Eq.~\ref{sigmasol2} corresponds to Eq.~2.9 in Cukier's analysis \cite{cukier1}. Note that in the operator notation,  $g_i$ multiplying a function implies a convolution over the $i^{\mathrm{th}}$ sphere surface: $g_i[\ldots]\equiv\int \mathrm{d} \Omega_i' g_i(\mathbf{r}_i,\mathbf{r}_i') [ \ldots ]_{\mathbf{r} = \mathbf{R}_i'}$.

  A useful ``scattering'' operator used in the following is
  \begin{align}
T_{i}(\mathbf{r},\mathbf{r}') &  \equiv \int \delta(\mathbf{r}-\mathbf{R}_i)g_i(\mathbf{r}_i,\mathbf{r}'_i)(1-\mathcal{Q}_i') \nonumber \\
& \qquad \qquad \qquad \qquad \times  \delta(\mathbf{r}'-\mathbf{R}_i') \, \mathrm{d} \Omega_i \, \mathrm{d} \Omega_i' .
\end{align}
 The operator $T_i$ describes the scattering of a nutrient off of the surface of the $i^{\mathrm{th}}$ cell. We now substitute Eq.~\ref{sigmasol2} into the $\tilde{\sigma}_i$ term in Eq.~\ref{exactsolution} and solve Eq.~\ref{exactsolution} for $\Psi$ by iteration: \begin{align}
\Psi & = G_W \left[ \sum_{i=1}^N T_i G_W\left(s-W \psi   + \sum_{j \neq i}  \tilde{\sigma}_j \right)-s+W \Psi \right] \nonumber \\
& =   \left[1- G_W \overline{T}+ \sum_{i,j\neq i} G_W T_i G_{W} T_{j }- \ldots \right] \nonumber \\ 
& \qquad \qquad \qquad \qquad \qquad \qquad \qquad \times G_W( W \Psi-s) \nonumber \\
& =- \left[1+G_{W}(\overline{T}-W)\right]^{-1}   G_Ws, \label{psiexpansion}
\end{align}
where  $\overline{T} \equiv \sum_iT_{i}$.  In Eq.~\ref{psiexpansion}, it is important to incorporate an exclusion condition for consecutive sums in the expansion of $[1+G_{W}(\overline{T}-W)]^{-1}$, e.g. $\overline{T} G_W \overline{T} =  \sum_{i,j\neq i}  T_i G_{W} T_{j }$ (see \cite{cukier1,cukier3} for more details).   
Note also that Eq.~\ref{effectivepropagator} (main text) implies that the ensemble averaged field $\psi = \langle \Psi \rangle$ in Fourier space satisfies
\begin{align}
& -D_0 q^2 \psi- W \psi =  \Sigma \psi - W \psi +s \nonumber \\
& \psi  = -G_W (\Sigma - W) \psi-G_W s \nonumber \\
& \psi = -\left[1+G_W( \Sigma-W) \right]^{-1} G_W s. \label{averagedsolution}
\end{align}
The field $\psi$ in Eq.~\ref{averagedsolution} must be equal to the ensemble average of Eq.~\ref{psiexpansion}.  We combine the two expressions and set $W = \Sigma$ to find a self-consistent equation for $\Sigma$.  After some algebraic manipulations (see Cukier's analysis \cite{cukier1} for details), the equation can be expanded in a series in  $\overline{T}$.    The expansion up to second order is
\begin{align}
\Sigma & \approx \left\langle \overline{T} \right\rangle- \sum_{i,j \neq i} \left\langle T_i G_{\Sigma} T_{j \neq i} \right\rangle+ \left\langle \overline{T}\right\rangle G_{\Sigma} \left\langle \overline{T} \right\rangle. \label{sigmaexpansion}
\end{align}
It makes sense to expand in $\overline{T}$ because this operator describes a single interaction of the nutrient with any of the cells.  The higher order terms in the expansion describe multiple scattering events, which we expect to be less probable.  We will calculate all three terms on the right hand side of Eq.~\ref{sigmaexpansion}, extending Cukier's analysis of just the first term \cite{cukier1}.

The various operators in Eq.~\ref{sigmaexpansion} are computed by moving to Fourier space and exploiting expansions in spherical harmonics.    For example,   $G_W(\Omega_i,\Omega_i')\equiv G_W(\mathbf{r}_i-\mathbf{r}
'_i)$ (given by Eq.~\ref{GWFT}) is expanded in  Fourier modes $e^{i \mathbf{q} \cdot (\mathbf{r}_i-\mathbf{r}_i')}$, which are then rewritten in terms of spherical harmonics via the spherical wave expansion of the plane wave \cite{jackson}. We find
\begin{align}
 G_W(\Omega_i,\Omega_i') =\sum_{\ell,m}  \gamma_{\ell} Y_{\ell m} (\Omega_{i}') Y^*_{\ell m} (\Omega_{i}) , \label{GWSH}
 \end{align}
 where $l=0,1,\ldots$,  $m=-\ell,-\ell+1,\ldots,\ell$, and
 \begin{align}
 \gamma_{\ell} \equiv \int_0^{\infty} \frac{2q^2\mathrm{d} q}{\pi} \,  \frac{  j_{\ell}(qa )^2 \,  }{D_0 q^2+ W(q)}. \label{gammaell}
 \end{align}
 The $j_{\ell}$ functions are the spherical Bessel functions of the first kind.   Cukier \cite{cukier1} describes this procedure in more detail.  Another important result  (Cukier's Eq.~A15) is 
\begin{align}
g_i(\mathbf{r}_i,\mathbf{r}_i') & = \sum_{\ell,m} [1-\gamma_{\ell}^{-1} \zeta_{\ell}]^{-1}\gamma_{\ell}^{-1}Y_{\ell m}(\Omega_i) Y_{\ell m}^*(\Omega_i') , \label{ktilde}
\end{align}
where 
\begin{align}
\zeta_{\ell} \equiv \frac{2a}{\pi \nu} \int_0^{\infty} \mathrm{d} q \,  \frac{ q^3 j'_{\ell}(qa )j_{\ell}(qa) \,  }{D_0 q^2+ W(q)}. \label{zetaell}
\end{align}

Now that we have an expression for $g_i$, it is possible to compute  $\left\langle\overline{T}\right\rangle$.  The bracket ensemble average  will require an averaging over all cell positions, so let us assume that a single cell center is distributed uniformly over the cluster volume $V$, so that $P(\mathbf{R}_i^0) = V^{-1}$ for all $i=1,\ldots,N$. After averaging, the operator $\langle \overline{T} \rangle$ acts on an arbitrary function $f(\mathbf{q})$ in Fourier space as follows:
\begin{align}
\left\langle \overline{T} \right\rangle f (\mathbf{q})& =c \int e^{-i \mathbf{q} \cdot \mathbf{r}_0}   g_0(\mathbf{r}_0,\mathbf{r}_0') (1-i\nu^{-1}\mathbf{r}_0' \cdot \mathbf{q}') \nonumber \\
& \quad  \times e^{i \mathbf{q}' \cdot \mathbf{r}_0'}  e^{i (\mathbf{q}'-\mathbf{q}) \cdot \mathbf{r}} f(\mathbf{q}')\, \mathrm{d} \Omega_0 \, \mathrm{d} \Omega_0' \, \frac{\mathrm{d} \mathbf{q}'}{(2 \pi)^3}  \, \mathrm{d} \mathbf{r}\nonumber \\  
& =4 \pi c \sum_{\ell}    \kappa_{\ell} (2 \ell+1) j_{\ell}(aq)   \nonumber \\
& \qquad \quad  \times  \left[j_{\ell}(aq)-\nu^{-1}aqj_{\ell}'(aq) \right]f(\mathbf{q}) \nonumber \\
& \equiv \mathcal{T}(\mathbf{q},\mathbf{q}) f(\mathbf{q}), \label{barTmomentumspace}
\end{align}
where $c=N/V$ is the concentration of cells and $\kappa_{\ell}~\equiv~[1-\gamma_{\ell}^{-1} \zeta_{\ell}]^{-1} \gamma_{\ell}^{-1}$.  In the last line of Eq.~\ref{barTmomentumspace}, we have introduced
  $\mathcal{T}(\mathbf{q},\mathbf{q}')$,  a convenient  scattering operator (averaged over the single cell center distribution $P(\mathbf{R}^0_i)=V^{-1}$).  In Fourier space, 
\begin{align}
\mathcal{T} (\mathbf{q},\mathbf{q}') & \equiv 4 \pi  \sum_{\ell} \kappa_{\ell} \left[j_{\ell}(aq')-\nu^{-1}a q'j_{\ell}'(aq') \right]  \nonumber \\
& \qquad \qquad  \times j_{\ell}(aq)(2\ell+1) P_{\ell}(\cos \theta_q), \label{frakToperator}
\end{align}
where $\theta_q$ is the angle between $\mathbf{q}$ and $\mathbf{q}'$ and $P_{\ell}$ are the Legendre polynomials.   Only the diagonal elements $\mathcal{T}(\mathbf{q},\mathbf{q})$ contribute to $\left\langle \overline{T} \right\rangle$.  The off-diagonal terms will contribute to the second order correction term $\sum_{i,j \neq i} \left\langle T_i G_{\Sigma} T_{j \neq i} \right\rangle$ in Eq.~\ref{sigmaexpansion}.

The second order correction term  requires the use of the pair distribution function $P(\mathbf{R}_i^0,\mathbf{R}_j^0)$ for the centers of pairs of cells $i$ and $j$.  We use the approximation Eq.~\ref{pairdistribution} for $P(\mathbf{R}_i^0,\mathbf{R}_j^0)$ and find\begin{align}
&\sum_{i,j \neq i}\left\langle T_i G_{\Sigma} T_{j \neq i} \right\rangle=
\frac{N(N-1)}{V^2}  \nonumber \\
& \qquad \times  \int \mathcal{T}(\mathbf{q},\mathbf{q}') \hat{G}_{\Sigma}(\mathbf{q}') \hat{\theta}(\mathbf{q}'-\mathbf{q})\mathcal{T}(\mathbf{q}',\mathbf{q}) \, \frac{\mathrm{d} \mathbf{q}'}{(2 \pi)^3}, \label{secondorderterm1}
\end{align}
where $\hat{\theta}(\mathbf{q})$ is the Fourier transform of $\theta(|\mathbf{r}|-2a)$:
\begin{align}
\hat{\theta}(\mathbf{q}) =-\,\frac{32\pi a^3 j_1(2aq)}{2aq} +(2 \pi)^3\delta^3(\mathbf{q}).
\end{align}   
We assume $N$ is large enough so that $N(N-1)\approx N^2$, and then substitute Eq.~\ref{secondorderterm1} into Eq.~\ref{sigmaexpansion} to find that to second order in scattering in the effective medium \begin{align}
&\Sigma(\mathbf{q}) =c\mathcal{T}(\mathbf{q},\mathbf{\mathbf{q}})   \nonumber \\
&  \quad+ \frac{4a^3c^2}{\pi^2}  \int \mathcal{T}(\mathbf{q},\mathbf{q}') \hat{G}_{\Sigma}(\mathbf{q}') \, \frac{j_1(2a|\mathbf{q}'-\mathbf{q}|)}{2a|\mathbf{q}'-\mathbf{q}|} \mathcal{T}(\mathbf{q}',\mathbf{q}) \, \mathrm{d} \mathbf{q}'. \label{secondordersolution}
\end{align}
In principle, Eq.~\ref{secondordersolution} could be solved for $\Sigma(\mathbf{q})$ numerically using an iterative procedure.  However, the second order terms involve an integral of  $\hat{G}_{\Sigma}(\mathbf{q}) = (D_0 q^2+\Sigma(q))^{-1}$  multiplied by various Bessel functions.  Due to the oscillatory nature of the Bessel functions, these kinds of integrals have poor convergence properties using standard numerical techniques and require special integration methods~\cite{besselintegration}.  To avoid these complications, we make a ``hydrodynamic'' approximation \cite{muthukumar,cukier1,cukier2} in which   $\Sigma(\mathbf{q})$ is approximated  by the first two terms of its Taylor expansion: $\Sigma(\mathbf{q}) \approx k+ \delta D q^2$.  

     The constants $\gamma_{\ell}$ and $\zeta_{\ell}$  have closed forms in the hydrodynamic approximation and involve integrations over spherical Bessel functions (tabulated in \cite{ryzhik}). We first note that\begin{align}
\gamma_{\ell} &= \frac{2}{D \pi a}\int_0^{\infty}   \frac{  j_{\ell}(x )^2 \,  }{ x^2+ \alpha^2} \,x^2 \,\mathrm{d} x=\frac{ \alpha i_{\ell} (\alpha) k_{\ell}(\alpha)}{D  a} , \label{gammahydro}
\end{align}
where $\alpha \equiv a/\xi$ is the ratio of the cell radius to the correlation length $\xi  = \sqrt{D/k}$ and $i_{\ell}(x)$,  $k_{\ell}(x)$ are the modified spherical Bessel functions of the first and second kinds, respectively.  Next, we have
\begin{align}
\zeta_{\ell} &=\frac{2}{\pi \nu D a} \int_0^{\infty}  \frac{  j'_{\ell}(x_+ )j_{\ell}(x) \,  }{x^2+ \alpha^2}  \, x^3 \, \mathrm{d} x  =\frac{\alpha^2i_{\ell}(\alpha)k'_{\ell}(\alpha)}{ \nu D a}, \label{zetahydro}
\end{align}
where the $x_+$ means we evaluate the derivative of the Bessel function at $x+\epsilon$ and let $\epsilon \rightarrow 0$.  Using Eq.~\ref{gammahydro} and Eq.~\ref{zetahydro} to evaluate $\mathcal{T}(\mathbf{q},\mathbf{q}')$ in Eq.~\ref{frakToperator} and substituting $\mathcal{T}(\mathbf{q},\mathbf{q}')$ into   Eq.~\ref{secondordersolution} (evaluated at $\mathbf{q}=0$) yields
\begin{align}
k & =  c\mathcal{T}(0,0) +\frac{8a^2c^2}{\pi D} \left[\frac{4 \pi Da\alpha \nu(1+\coth \alpha)   }{1+ \alpha +\nu} \right]^2\nonumber  \\
& \qquad   \times   \int_0^{\infty} \frac{ j_{0}(x)j_1(2x)\left[j_{0}(x)-\nu^{-1}xj_{0}'(x) \right]}{ x^2+ \alpha^2 } \,x \, \mathrm{d}x 
, \label{solutionforalpha}
\end{align}
where the integral over the product of three spherical Bessel functions is performed by writing the Bessel functions in terms of elementary functions such as powers and exponentials and then using contour integration techniques. The final result reduces to the implicit equation for $\xi$, given in the main text as Eq.~\ref{finalxisolution}.

The diffusion coefficient $D$ can be calculated by looking at the second order term in the Taylor expansion of Eq.~\ref{secondordersolution} around $\mathbf{q}=0$.  Note that for the purposes of determining the effects on nutrient shielding in a cell cluster, the relevant quantity is the screening length $\xi$.  A lengthy calculation of the change in the diffusion coefficient $\delta D = D - D_0$ yields
\begin{align}
  \frac{\delta D}{D}  & =  \frac{c}{2D}   \left.\frac{d^2}{dq^2}  \left[  \mathcal{T}(\mathbf{q},\mathbf{\mathbf{q}}) \right] \right|_{\mathbf{q=0}} \nonumber \\
& +\frac{a^3c^2}{\pi^2 D} \left.\frac{d^2}{dq^2}  \left[   \int \mathrm{d} \mathbf{q}'  \, \mathcal{T}(\mathbf{q},\mathbf{q}') \hat{G}_{\Sigma}(\mathbf{q}') \right. \right. \nonumber \\
& \qquad \qquad \qquad  \qquad \quad \left. \left. {}\times \frac{j_1(2a|\mathbf{q}'-\mathbf{q}|)}{a|\mathbf{q}'-\mathbf{q}|} \mathcal{T}(\mathbf{q}',\mathbf{q})   \right] \right|_{\mathbf{q=0}}
\nonumber \\
 & = \frac{ (1-\nu)( \tilde{\kappa}_0- \tilde{\kappa}_1)}{\nu} \left[1+ \frac{6  \tilde{\kappa}_0}{\alpha^2} \, \phi \right]\phi-  \frac{9 \tilde{\kappa}_0^2}{ \alpha^4} \, \phi^2   \nonumber \\
& \, \, {}+36 \left\{   \tilde{\kappa}_0i_0(\alpha)\left[ \left(  \frac{1}{\alpha^2}+1- \frac{1}{3 \nu} \right)  k_1(2\alpha)+  \frac{2k_0(2\alpha)  }{3 \alpha  }    \right]  \right. \nonumber \\
& \left.    {}+\frac{2(1-\nu)k_2(2\alpha) \tilde{\kappa}_1 i_1(\alpha)}{3 \nu}\right\}   \tilde{\kappa}_0\left[  i_0(\alpha)- \frac{\alpha i'_0(\alpha) }{\nu}\right]  \phi^2 \nonumber \\
&\, \, \,  {}+\frac{12 \tilde{\kappa}_1}{ \nu} \left[ \tilde{\kappa}_1 (\nu-1)k_1(2\alpha) i_1(\alpha)  -2 \nu   k_2(2\alpha) \tilde{\kappa}_0  i_0(\alpha)\right] \nonumber \\
& \qquad \qquad \qquad \times  \left[ i_1(\alpha)-\frac{\alpha i_1'(\alpha)}{\nu} \,   \right]  \phi^2 , \label{solutionfordiff}
\end{align}
where $\alpha \equiv a/\xi $ is given by the self-consistent solution to Eq.~\ref{finalxisolution} and   $ \left[ \tilde{\kappa}_{\ell}(\alpha) \right]^{-1} \equiv \alpha i_{\ell}( \alpha )   \left[ k_{\ell}( \alpha )-\frac{ \alpha k_{\ell}'( \alpha )}{\nu  } \right]$ for $\ell=0,1$.
\bibliographystyle{biophysj}
\bibliography{shielding}

\end{document}